\documentclass[twocolumn,epjc3]{svjour3}          

\RequirePackage[T1]{fontenc}

\smartqed  

\RequirePackage{graphicx}
\RequirePackage{mathptmx}      
\RequirePackage{flushend}
\RequirePackage[numbers,sort&compress]{natbib}
\RequirePackage[colorlinks,citecolor=blue,urlcolor=blue,linkcolor=blue]{hyperref}

\RequirePackage{hyphenat}

\RequirePackage{amsmath}
\RequirePackage{framed}

\RequirePackage{placeins}
\RequirePackage{mathrsfs}
\RequirePackage{wasysym}
\RequirePackage{verbatim}
\RequirePackage{booktabs}
\RequirePackage[tight]{subfigure}
\RequirePackage{color}
\RequirePackage{graphicx}
\RequirePackage[utf8]{inputenc}

\newcommand{\be}{\begin{equation}}
\newcommand{\ee}{\end{equation}}
\newcommand{\bea}{\begin{eqnarray}}
\newcommand{\eea}{\end{eqnarray}}

\newcommand{\df}{\dfrac}

\newcommand\simgt{\stackrel{>}{{}_\sim}}
\newcommand{\gsim}{\lower.7ex\hbox{$\;\stackrel{\textstyle>}{\sim}\;$}}
\newcommand{\lsim}{\lower.7ex\hbox{$\;\stackrel{\textstyle<}{\sim}\;$}}

\newcommand\co{{\mathcal O}}

\newcommand\Mm{M_{\rm mess}}

\journalname{Eur. Phys. J. C}
\begin{document}

\title{Reducing the Fine-Tuning of Gauge-Mediated SUSY Breaking}

\author{J. Alberto Casas\thanksref{e1,addr1}
        \and
        Jes\'us M. Moreno\thanksref{e2,addr1} 
        \and
        Sandra Robles\thanksref{e3,addr1,addr2}       
        \and
        Krzysztof Rolbiecki\thanksref{e4,addr1,addr3}            
}

\thankstext{e1}{e-mail: alberto.casas@uam.es}
\thankstext{e2}{e-mail: jesus.moreno@csic.es}
\thankstext{e3}{e-mail: sandra.robles@uam.es}
\thankstext{e4}{e-mail: krzysztof.rolbiecki@desy.de}

\institute{Instituto de F\'isica Te\'orica, IFT-UAM/CSIC, Universidad Aut\'onoma de Madrid, 
        Cantoblanco, 28049 Madrid, Spain \label{addr1}
          \and
         Departamento de F\'{\i}sica Te\'{o}rica, Universidad Aut\'{o}noma de Madrid,  
        Cantoblanco, 28049 Madrid, Spain \label{addr2}
          \and
          Faculty of Physics, University of Warsaw, 02093 Warsaw, Poland\label{addr3}
}

\date{Received: date / Accepted: date}    


\maketitle           

\begin{abstract}
Despite their appealing features, models with gauge\hyp{}mediated supersymmetry breaking (GMSB)
typically present a high degree of fine\hyp{}tuning, due to the initial absence of the top trilinear scalar couplings, $A_t=0$. 
In this paper, we carefully evaluate such a tuning, showing that is worse than per mil in the minimal model. 
Then, we examine some existing proposals to generate $A_t\neq 0$ term in this context. 
We find that,  although the stops can be made lighter, usually the tuning does not improve (it may be even worse), with some exceptions, which involve the generation of $A_t$ at one loop or tree level. We examine both possibilities and propose a conceptually simplified version of the latter; which is arguably the optimum GMSB setup (with minimal matter content), concerning the fine\hyp{}tuning issue. 
The resulting fine\hyp{}tuning is better than one per mil, still severe but similar to other minimal supersymmetric standard model constructions.
We also explore the so\hyp{}called ``little $A_t^2/m^2$ problem'', i.e.\ the fact that a large $A_t$\hyp{}term is normally accompanied by a similar or larger sfermion mass, which typically implies an increase in the fine\hyp{}tuning. Finally, we find the version of GMSB for which this ratio is optimized, which, nevertheless, does not minimize the fine\hyp{}tuning. 
\end{abstract}


\section{Introduction\label{sec:1}}

Models with gauge-mediated supersymmetry breaking (GMSB) \cite{Dine:1981za,Dimopoulos:1981au,Dine:1981gu,Dine:1982qj,Dine:1982zb,AlvarezGaume:1981wy,Nappi:1982hm,Dimopoulos:1982gm,Dine:1993yw,Dine:1994vc,Dine:1995ag}, have become one of the most popular supersymmetric scenarios. In these models the breakdown of supersymmetry (SUSY) takes place in a hidden sector and is radiatively transmitted to the visible sector via heavy particles (messengers) that are charged under the standard gauge interactions. 
The main merit of GMSB models is that they automatically imply universality of soft terms (associated to fields with the same quantum numbers), thus avoiding dangerous flavour-changing neutral currents (FCNC) effects. On the other hand, these models typically present a high degree of fine-tuning, a problem which is accentuated by the rather high Higgs mass and by the initial absence of (stop) scalar trilinear coupling. In this paper, we carefully compute this fine-tuning and explore the possibilities to reduce it as much as possible, keeping the minimal matter content.

Let us briefly review the formulation of GMSB models. One starts with a set of messenger superfields coupled to the superfield $X$ which breaks SUSY in the hidden sector, thanks to a non-vanishing VEV of its auxiliary component, $\langle F_X\rangle\neq0$. Typically, the scalar component of $X$ gets a VEV as well, contributing to the masses of the messengers. Schematically, the relevant superpotential reads
\be
W_{\rm mess} = kX\bar\Phi\Phi + \hat M_{\rm mess} \bar \Phi\Phi \ ,
\label{Wmess}
\ee 
where $\Phi$ and $\bar\Phi$ collectively denote the messenger superfields, $k$ is a dimensionless coupling and $\hat M_{\rm mess}$ is a messenger mass term. In general, there can be different couplings and masses for the various messengers, though usually they are taken universal for simplicity. Then, without loss of generality, one can re-define the scalar component of $X$ either to make $\hat M_{\rm mess}=0$ or $\langle X\rangle=0$. 
The masses of the fermionic components of the messengers are simply $M_{\rm mess}= \hat M_{\rm mess}+k\langle X\rangle$, while the masses of the scalar partners arise from the mass-squared matrix
\bea
\left(
\begin{array}{cc}
\Mm^2\ \ \ \ \ \ \ & (kF_X)^\dagger \\
(kF_X)\ \ \ \ \ \ \  & \Mm^2 \\
\end{array} \right) \ .
\eea
Consequently, the requirement of positive masses demands
\be
\label{xlimit}
x<1 \ ,
\ee
where
\be
\label{x}
x\equiv \frac{\Lambda}{M_{\rm mess}}\ , \quad \Lambda\equiv\frac{kF_X}{M_{\rm mess}} \ .
\ee

If the messengers form complete $SU(5)$ representations, then gauge unification is preserved. Hence, a usual (and somehow minimal) choice is that the messenger sector consists of $N_5$ copies of fundamental representations, $5+\bar 5$. 
With this minimal content, the gauginos and sfermions of the minimal supersymmetric standard model (MSSM) get masses at one loop and two-loops respectively, namely 
\cite{Dimopoulos:1996gy,Martin:1996zb,Poppitz:1996xw}:
\begin{eqnarray}
M_i &=& \df{\alpha_i}{4\pi}\Lambda N_5\left[1+{\cal O}(x^2)\right] \ ,
\label{MiGM} \\
m^2_{\tilde f} &=& 2\Lambda^2 N_5 \sum_{i=1}^3 C_i^{\tilde f} \left(\df{\alpha_i}{4\pi}\right)^2 \left[1+{\cal O}(x^2)\right]\ .
\label{MsGM}
\end{eqnarray}

Here $\alpha_i=g_i^2/4\pi$ stand for the usual gauge couplings, $C_i$ are the corresponding quadratic Casimir (see \ref{MGMSBappendix} for further details). The above expressions are to be understood at the high scale, $M_{\rm HE}$, where the effects of SUSY breaking are transmitted to the observable sector, which coincides with the messenger mass, $M_{\rm HE}=M_{\rm mess}$. Altogether the {\em minimal} GMSB scenario has only four independent parameters,
\be
\label{parGMSB}
\{\Lambda, M_{\rm mess}, \mu, B\} 
\ee
(plus the discrete ${\cal O}(1)$ number $N_5$), in contrast with the 5 parameters of the constrained MSSM: $\{m_0, M_{1/2}, A, \mu, B\}$. Hence, the GMSB is a highly predictive and well-motivated MSSM, and thus with extremely interesting phenomenology. In this sense, a distinctive feature of GMSB models is that, unlike the constrained MSSM, the soft masses are different for particles with different quantum numbers, although they are independent of the family. This partial universality is enough to avoid dangerous FCNC effects, which is an important success of GMSB. 

On the other hand, there is no clear mechanism to generate neither a $\mu$-term for the two Higgses in the superpotential ($W\supset \mu H_u H_d$), nor the corresponding soft bilinear scalar coupling, $B$. A usual procedure is to assume that $\mu$ and $B$ have appropriate values at low energy in order to produce the required VEVs for the two Higgses, $\langle H_u\rangle^2 + \langle H_d\rangle^2=\langle H_{\rm SM}\rangle^2$, and a reasonable value of $\tan\beta\equiv \langle H_u\rangle/\langle H_d\rangle$. Incidentally, this is exactly the strategy followed in the constrained MSSM.

One of the most problematic aspects  of the GMSB scenario is the initial absence of trilinear scalar couplings, $A_i=0$, in particular the one associated with the top, $A_t$.
Although a non-vanishing $A_t$ is generated along the renormalization group (RG) running from high to low energy, its final value is rather small. The consequence is that the threshold correction to the Higgs mass, $m_h$, is far from its maximal value, and thus the stop masses must be quite large in order to generate sizeable radiative corrections to $m_h$, able to reconcile its value with the experimental one. Such large stop masses (around 10 TeV) imply in turn a severe fine-tuning in order to get the right electroweak (EW) breaking scale. The reason is that, along the RG running, the soft masses of the Higgses (in particular $m_{H_u}^2$) receive important contributions proportional to the stop masses. Then a tuning of parameters (essentially between $m_{H_u}^2$ and $\mu^2$) is necessary to get the correct expectation values of the Higgses. Typically such large stops lead to fine-tunings of one per mil or per ten thousand \cite{Casas:2014eca}.

This problem has been addressed in the literature following different strategies \cite{Dine:1996xk,Giudice:1997ni,Chacko:2001km,Chacko:2002et,Shadmi:2011hs,Evans:2011bea, Jelinski:2011xe, Evans:2012hg,Kang:2012ra,Craig:2012xp,Albaid:2012qk,Abdullah:2012tq, Perez:2012mj, Endo:2012rd,Kim:2012vz,Byakti:2013ti,Craig:2013wga,Evans:2013kxa,Calibbi:2013mka,Jelinski:2013kta,Galon:2013jba,Fischler:2013tva,Knapen:2013zla,Ding:2013pya,Calibbi:2014yha,Basirnia:2015vga}. Keeping a minimal matter content, the only way out is to devise some mechanism able to generate the desired $A$-terms ab initio. This requires non-trivial couplings between the messengers and the MSSM superfields in the superpotential. 
The most studied scenarios involve the generation of  $A$-terms through loops. This idea was first considered in ref.~\cite{Dine:1996xk} and further developed in ref.~\cite{Giudice:1997ni} and in many other papers
\cite{Chacko:2001km, Chacko:2002et, Shadmi:2011hs,
Evans:2011bea, Jelinski:2011xe, Evans:2012hg, Craig:2012xp,Albaid:2012qk, Abdullah:2012tq, Perez:2012mj, Endo:2012rd,
Evans:2013kxa,Calibbi:2013mka}. In ref.~\cite{Evans:2013kxa}, Evans and Shih performed an extensive survey of this type of models, finding out the most favourable ones for the fine-tuning. 
Also, in ref.~\cite{Calibbi:2013mka}, Calibbi et al. studied in depth a model of this kind, showing explicitly how a maximal $A_t$ can be generated, allowing for much lighter stops. 

Later, a mechanism for tree\hyp{}level generation of an $A_t$\hyp{}term has been explored in ref.~\cite{Basirnia:2015vga}, where the authors stress the so-called  ``little $A_t^2/m^2$ problem'' \cite{Craig:2012xp}, i.e.\ the fact that a large $A_t$\hyp{}term is normally accompanied by a similar or larger sfermion mass-squared, which typically implies an increase in the fine-tuning. 

In this paper, we re\hyp{}visit the computation and the prospects of the fine\hyp{}tuning associated with GMSB models and propose a simple scenario, which alleviates this problem as much as it is possible (at least playing with minimal matter content). 
In section~\ref{sec:2}, we expound the strategy for the computation of the fine\hyp{}tuning in the MSSM, particularizing to the GMSB scenario. We also comment on the importance of a reliable computation of the Higgs mass, especially when the stops are heavy (which is the usual case in GMSB). In this sense, we use the most recent codes for the Higgs mass computation, showing that previous analyses underestimated the fine\hyp{}tuning of GMSB. In section~\ref{sec:MGMSB}, we compute the fine\hyp{}tuning of the minimal GMSB set up, showing that it is a few per ten thousand. Section~\ref{sec:4} is devoted to models with radiatively generated $A$\hyp{}terms. We refine the fine\hyp{}tuning calculation for the most favourable case, according to Evans and Shih \cite{Evans:2013kxa}, and compute it for the scenario proposed by Calibbi et al. \cite{Calibbi:2013mka}. We show that, in the latter case, even though the stop masses become smaller than in the minimal GMSB, the fine\hyp{}tuning does not improve; actually, it gets worse. In section~\ref{sec:5}, we consider the tree\hyp{}level generation of $A$\hyp{}terms, in the spirit of ref.~\cite{Basirnia:2015vga}. We explore this scenario, simplifying it to some extent and looking for the version that optimizes the fine\hyp{}tuning (which does not necessarily coincides with the one that minimizes the little $A_t^2/m^2$ problem). In the best case scenario, the fine\hyp{}tuning can be better than one per mil. This is still a severe fine\hyp{}tuning, but much milder than other versions of gauge\hyp{}mediation (at least with minimal observable matter content), and of the same order as in other MSSMs. Finally, in section~\ref{sec:6} we present our conclusions.

\section{Computing the electroweak fine-tuning\label{sec:2}}

\subsection{The fine-tuning of the MSSM}

Let us start considering the origin and measure of the electroweak fine-tuning in the MSSM, as the GMSB scenario is a particular case of it. In the MSSM, the vacuum expectation value of the Higgs, $v^2/2 = |\langle H_u\rangle|^2 +|\langle H_d\rangle|^2$, is given at tree level by the minimization relation
\bea
\label{mu}
-\frac{1}{8}(g^2+g'^2)v^2 = -\frac{M_Z^2}{2}=\mu^2-\frac{m_{H_d}^2 - m_{H_u}^2\tan^2\beta}{\tan^2\beta-1} \ .
\eea
As is well known, the value of $\tan\beta$ must be rather large, so that the tree-level Higgs mass,  $(m_h^2)_{\rm tree-level}=M_Z^2 \cos^22\beta$, is as large as possible, $\simeq M_Z^2$; otherwise, the radiative corrections needed to reconcile the Higgs mass with its experimental value, would imply gigantic stop masses  and thus an extremely fine-tuned scenario. Then eq.~(\ref{mu}) gets simplified, 
\bea
\label{min}
-\frac{1}{8}(g^2+g'^2)v^2 = -\frac{M_Z^2}{2}=\mu^2+m_{H_u}^2\ .
\eea
The absolute values of the two terms on the  r.h.s.\ are typically much larger than $M_Z^2$, hence the potential fine-tuning associated to the electroweak breaking. The radiative corrections to the Higgs potential somewhat reduce the fine-tuning due to the running of the effective quartic coupling of the SM-like Higgs from its initial value at the SUSY threshold,\footnote{A convenient choice of the SUSY-threshold is the average stop mass, since the one-loop correction to the Higgs potential is dominated by the stop contribution.} $\lambda(Q_{\rm threshold})=\frac{1}{8}(g^2+g'^2)$, until its final value at the electroweak scale, $\lambda(Q_{EW})$. Essentially, this is equivalent to replace $M_Z^2 \rightarrow m_h^2$ in eq.~\eqref{min} above (for more details see ref.~\cite{Casas:2014eca}), i.e.
\bea
\label{minh}
-\frac{m_h^2}{2}=\mu^2+m_{H_u}^2\ ,
\eea
which is the expression from which we evaluate the electroweak fine-tuning in the MSSM. We emphasize here that in this expression $\mu^2$ and $m_{H_u}^2$ are to be understood at low energy.

In order to quantify that fine-tuning a common practice is to use the parametrization first proposed by Ellis et al.~\cite{Ellis:1986yg} and Barbieri and Giudice~\cite{Barbieri:1987fn}, which in our case reads
\bea
\label{BG}
\frac{\partial m_h^2}{\partial \theta_i} = \Delta_{\theta_i}\frac{ m_h^2}{\theta_i}\ , \ \ \ \ \ \Delta\equiv {\rm Max}\ \left|\Delta_{\theta_i}\right|\ ,
\eea
where $\theta_i$ is an independent parameter that defines the model under consideration and $\Delta_{\theta_i}$ is the fine-tuning parameter associated to it. Very often in the literature (see e.g.~\cite{Feng:2013pwa}), the initial (high-energy) values of the soft terms and the $\mu$ parameter are considered as the independent $\theta_{i}$ parameters in the previous expression. However, for specific scenarios of SUSY breaking and transmission to the observable sector, the initial parameters are those that define the scenario and hence determine the soft terms as a by-product. We will discuss this point soon for the specific case of the GMSB framework.

The value of $\Delta$ can be interpreted as the inverse of the $p$-value to get $m_h^2$ from eq.~(\ref{minh}) equal or smaller than
the experimental $m_h^2$. Note here that if $\theta$ is the parameter that  gives the maximum $\Delta$-parameter and $\delta \theta$ represents the $\theta$-interval for which $m_h^2\lsim (m_h^{\rm exp})^2$, then
\begin{eqnarray}
\label{toy2}
p{\rm -value} \simeq \left|\frac{\delta \theta}{\theta_0}\right| \equiv \Delta^{-1}\ .
\end{eqnarray}
where we have expanded $m_h(\theta)^2$ to first order; for more details on the statistical meaning of $\Delta$ see refs.~\cite{Ciafaloni:1996zh,Casas:2014eca}.

\subsection{Application to GMSB models}\label{tunGMSB}

The low-energy (LE) values of $\mu^2$ and $m_{H_u}^2$ entering eq.~(\ref{minh}) are related to the high-energy (HE) values of {\em all} the soft masses and $\mu$ through the RG-equations. Fortunately, dimensional and analytical consistency dictates the form of the dependence,
\begin{eqnarray}
\label{mHu_gen_fit}
m_{H_u}^2({\rm LE})&=&
c_{M_3^2}M_3^2 +c_{M_2^2}M_2^2 +c_{M_1^2}M_1^2 + c_{A_t^2}A_t^2 \nonumber\\
&&+c_{A_tM_3}A_tM_3 +c_{M_3M_2}M_3M_2+ \cdots 
\nonumber\\
&&+c_{m_{H_u}^2}m_{H_u}^2 +c_{m_{Q_3}^2} m_{Q_3}^2 +c_{m_{U_3}^2}m_{U_3}^2+\cdots\ ,
\\
\mu({\rm LE})&=&c_\mu \mu \ ,
\label{mu_gen_fit}
\end{eqnarray}
where $M_i$ are the $SU(3)\times SU(2)\times U(1)_Y$ gaugino masses; $A_t$ is the top trilinear scalar coupling; and $m_{H_u}, m_{Q_3}, m_{U_3}$ are the masses of the $H_u$ Higgs doublet, the third-generation squark doublet and the stop singlet, respectively, all of them understood at the HE scale. The numerical coefficients, $c_{M_3^2},$ $c_{M_2^2},\ldots$ are obtained by fitting the result of the numerical integration of the RGEs to eqs.~\eqref{mHu_gen_fit} and \eqref{mu_gen_fit}, a task that was carefully performed in ref.~\cite{Casas:2014eca}.

Nevertheless, the independent parameters of GMSB are not the HE soft parameters, but $\Lambda$, $\mu$, $M_{\rm mess}$ and $B$, as in eq.~\eqref{parGMSB}. In particular, for the simplest GMSB, the gaugino and scalar masses are given by eqs.~\eqref{MiGM} and \eqref{MsGM}, while $A_i=0$. 
Therefore, neglecting the higher-order corrections in $x$ in eqs.~\eqref{MiGM} and \eqref{MsGM}, i.e.\ for $x\ll 1$, the r.h.s.\ of eq.~(\ref{mHu_gen_fit}) is proportional to $\Lambda^2$, as well as $\mu({\rm LE})$ is proportional to its initial value at HE.  Plugging these expressions in eqs.~\eqref{minh} and  \eqref{BG} we find that the fine-tuning in $\Lambda$ and $\mu$ simply read
\begin{eqnarray}
\label{BGLambda}
\left|\Delta_{\Lambda}\right|=\left|\frac{\Lambda}{m_h^2}\frac{\partial m_h^2}{\partial \Lambda}\right| 
=4 \left|\df{m_{H_u}^2({\rm LE})}{m_h^2} \right| \ ,
 \end{eqnarray} 
\begin{eqnarray}
\label{BGmu}
\left|\Delta_{\mu}\right|=\left|\frac{\mu}{m_h^2}\frac{\partial m_h^2}{\partial \mu}\right| 
=4 \df{\mu^2({\rm LE})}{m_h^2}   \ .
 \end{eqnarray} 
Since $\left|m_{H_u}^2({\rm LE})\right|\simeq \left|\mu^2({\rm LE})\right|$, the fine-tuning associated to $\Lambda$ and $\mu$ are almost exactly the same. Actually, they are somehow redundant since the value of $m_h^2$ arises as a cancellation between both quantities, eq.~(\ref{minh}). 

Of course, for a particular value of $\Lambda$, the corresponding $m_{H_u}^2({\rm LE})$ depends on the initial HE scale at which eqs.~\eqref{MiGM} and \eqref{MsGM} should be evaluated. Therefore, the EW fine-tuning depends on $\Mm$.  Actually, by continuity one expects a value of $\Mm$ for which $m_{H_u}^2({\rm LE})=0$, since for large $\Mm$, say $\Mm=M_X$, $m_{H_u}^2({\rm LE})$ is negative, while for $\Mm=M_{{\rm LE}}$ it is positive. So, there is a particular choice of $\Mm$ between these two scales for which $m_{H_u}^2({\rm LE})=0$, and therefore the fine-tuning disappears! In other words, for some clever choice of the high-energy scale the simplest GMSB scenario presents a global focus point. We will see soon which scale is that. But, in any case, notice that this is not the end of the story. $\Mm$ is an independent parameter itself, so if we allow ourselves to choose it at convenience there is a fine-tuning parameter associated with $\Mm$,
\begin{eqnarray}
\label{BGMmess}
\left|\Delta_{\Mm}\right|=\left|\frac{\Mm}{m_h^2}\frac{\partial m_h^2}{\partial \Mm}\right| \simeq 2\ \left|\frac{\Mm}{m_h^2}\frac{\partial m_{H_u}^2({\rm LE})}{\partial \Mm}\right| 
 \ .
 \end{eqnarray} 
This fine-tuning is normally smaller than the one associated to $\Lambda$, since the dependence of $m_{H_u}^2({\rm LE})$ on ${\Mm}$ is only logarithmic. 

Finally, if one goes beyond the minimal GMSB model, e.g.\ by including non-trivial couplings between the messengers and the chiral fields in the superpotential, as mentioned in the Introduction, then there are additional independent parameters (the values of those couplings), whose associated fine-tuning should be computed and taken into account in eq.~(\ref{BG}). All these issues will be illustrated in the following sections.

\subsection{The Higgs mass issue}\label{sect:mh}

As is well known, radiative corrections to the Higgs mass are needed in the MSSM in order to reconcile it with the experimental value. 
A simplified expression of such corrections, obtained at the leading-log approximation \cite{Casas:1994us,Carena:1995bx,Haber:1996fp} is
\begin{equation}\label{eq:der2}
\delta m_{h}^{2}=\frac{3 G_{F}}{\sqrt 2 \pi^{2}}m_{t}^{4}\left(\log\left(\frac{\overline m_{\tilde t}^{2}}{m_{t}^{2}}\right)+\frac{X_{t}^{2}}{\overline m_{\tilde t}^{2}}\left(1-\frac{X_{t}^{2}}{12\overline m_{\tilde t}^{2}}\right)\right)\ +\ \cdots\ ,
\end{equation}
with $\overline m_{\tilde t}$ the average stop mass and $X_{t}=A_{t}-\mu \cot\beta$. 
The $X_t$-contribution arises from the threshold corrections to the quartic coupling at the stop scale. 
At this level, the threshold correction is maximized for $X_{t}=\pm\sqrt 6 \overline m_{\tilde t}$. 

In order to obtain a reliable expression for the Higgs mass, especially when the stops are heavier than 1 TeV, higher-order corrections are crucial. 
There are in the literature several codes that cope with this problem. Among the most recent ones are the last versions of \texttt{FeynHiggs}~\cite{Heinemeyer:1998yj,Heinemeyer:1998np,Degrassi:2002fi,Frank:2006yh,Hahn:2013ria} 
and \texttt{SusyHD}~\cite{Vega:2015fna}. It turns out that, typically, previous codes overestimated the Higgs mass in the large-stop-mass regime. 
This is quite relevant for the computation of the GMSB fine-tuning. In these models, the absence of an initial $A_t$ soft term implies that the threshold correction is far from maximal, hence $m_h\simeq 125$ GeV requires large stop masses. The fact that the latter were underestimated in previous codes implies that the required value of $\Lambda$, and thus the fine-tuning, was also underestimated.

\begin{figure*}[ht]
\centering 
\includegraphics[width=0.8\linewidth]{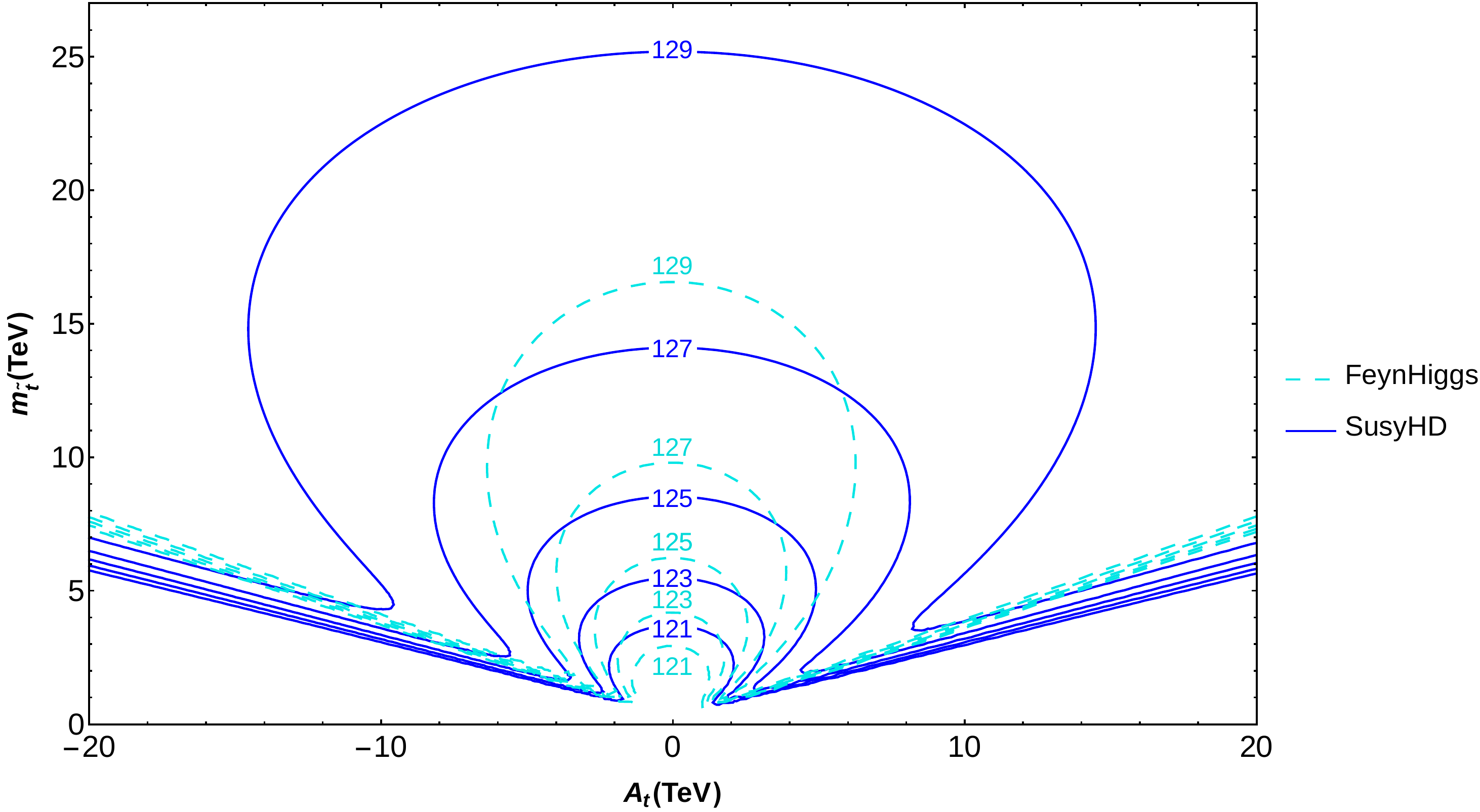}
\caption{Contour lines of constant $m_h$ in the $\overline{m}_{\tilde{t}}$--$A_t$ plane. 
The dashed cyan lines and the solid blue lines correspond to the Higgs mass calculated with \texttt{FeynHiggs} and \texttt{SusyHD}, respectively.
}
\label{fig:mhSUSYHD}
\end{figure*}

In order to illustrate the Higgs mass dependence on the (averaged) stop mass and $A_t$ at the LE scale,
we show in figure~\ref{fig:mhSUSYHD} contour lines of constant $m_h$ in the $\overline{m}_{\tilde{t}}$--$A_t$ plane. 
We have calculated the Higgs mass with \texttt{FeynHiggs~2.11.3}~\cite{Heinemeyer:1998yj,Heinemeyer:1998np,Degrassi:2002fi,Frank:2006yh,Hahn:2013ria} (dashed cyan lines) 
and \texttt{SusyHD~1.0.2}~\cite{Vega:2015fna} (solid blue lines) with parameters in the OS-scheme, taking the soft stop masses as degenerate for simplicity, $\mu=200$ GeV and $\tan\beta=10$. Note that for a moderately large value of $\tan\beta$, as usual, $X_t\simeq A_t({\rm LE})$.

For a given value of the Higgs mass, the minimum stop mass occurs for the two values of $A_t({\rm LE})$ that maximize the threshold correction, $A_t({\rm LE})-\mu\cot\beta\simeq \pm 2\overline{m}_{\tilde{t}}$ (note that this value slightly departs from the previous leading-order one, $\pm \sqrt 6 \overline m_{\tilde t}$). 
As long as $A_t({\rm LE})$ departs from the maximizing value, larger stop masses are required to reproduce the Higgs mass. Typically, for the same stop mass, the \texttt{FeynHiggs} result for $m_h$ is $\sim 2$ GeV larger than the \texttt{SusyHD} one. This has a non-negligible impact in the calculation of the fine-tuning. In the next sections, we present our results using both codes.

\section{The fine-tuning of the minimal GMSB}\label{sec:3}
\label{sec:MGMSB}
\begin{figure}[ht]
\centering 
\includegraphics[width=1.0\linewidth]{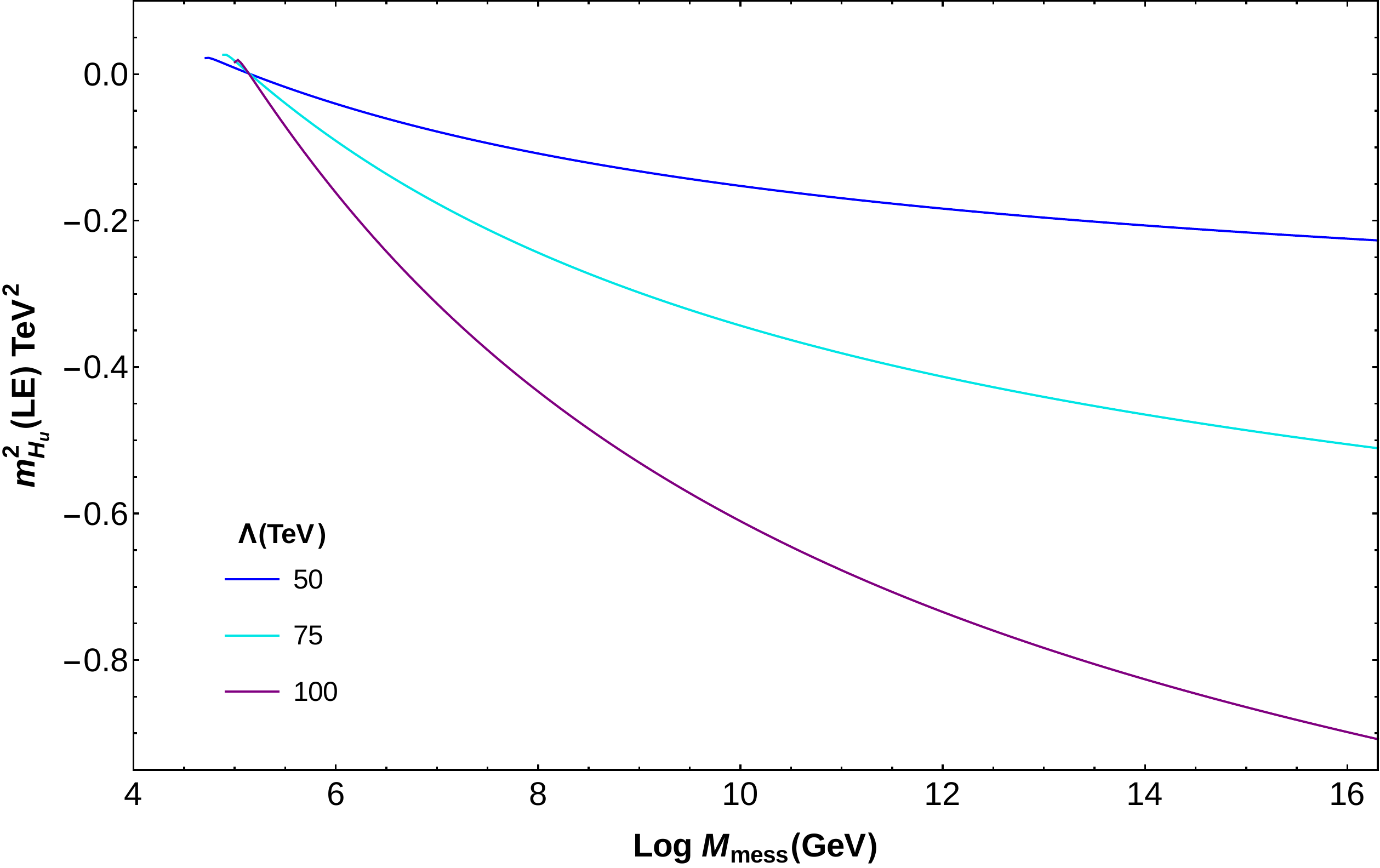}
\caption{$m_{H_u}^2({\rm LE})$ vs. $\Mm$ for $N_5=3$ and different choices of $\Lambda$. Note the focus-point behaviour around $\Mm \simeq 10^5$ GeV. 
}
\label{fig:focuspoint}
\end{figure}

The minimal GMSB scenario, as it was defined in the Introduction (see eq.~(\ref{parGMSB})), has only four independent parameters, $\{\Lambda, M_{\rm mess}, \mu, B\}$. The tuning associated with $\Lambda$ is given by eq.~(\ref{BGLambda}), which is equivalent to that of $\mu$, eq.~(\ref{BGmu}). As discussed in section~\ref{tunGMSB}, one expects some value of $\Mm$ for which $m_{H_u}^2({\rm LE})=0$. 
This is illustrated in figure~\ref{fig:focuspoint}, which shows $m_{H_u}^2({\rm LE})$ vs.\ $\Mm$ for fixed $\Lambda$ and $N_5=3$. 
Since all gaugino (sfermion) masses are proportional to $\Lambda$ ($\Lambda^2$), then $m_{H_u}^2({\rm LE})\propto \Lambda^2$, so the corresponding curves for different choices of $\Lambda$ are easy to draw. The important point is that for $\Mm \simeq 10^5$ GeV one gets $m_{H_u}^2({\rm LE})=0$, independently of the value of $\Lambda$, exhibiting a global fixed-point. Thus, for that choice of $\Mm$, the electroweak fine-tuning associated with $\Lambda$ vanishes!

\begin{figure*}[ht]
\centering 
\includegraphics[width=.49\linewidth]{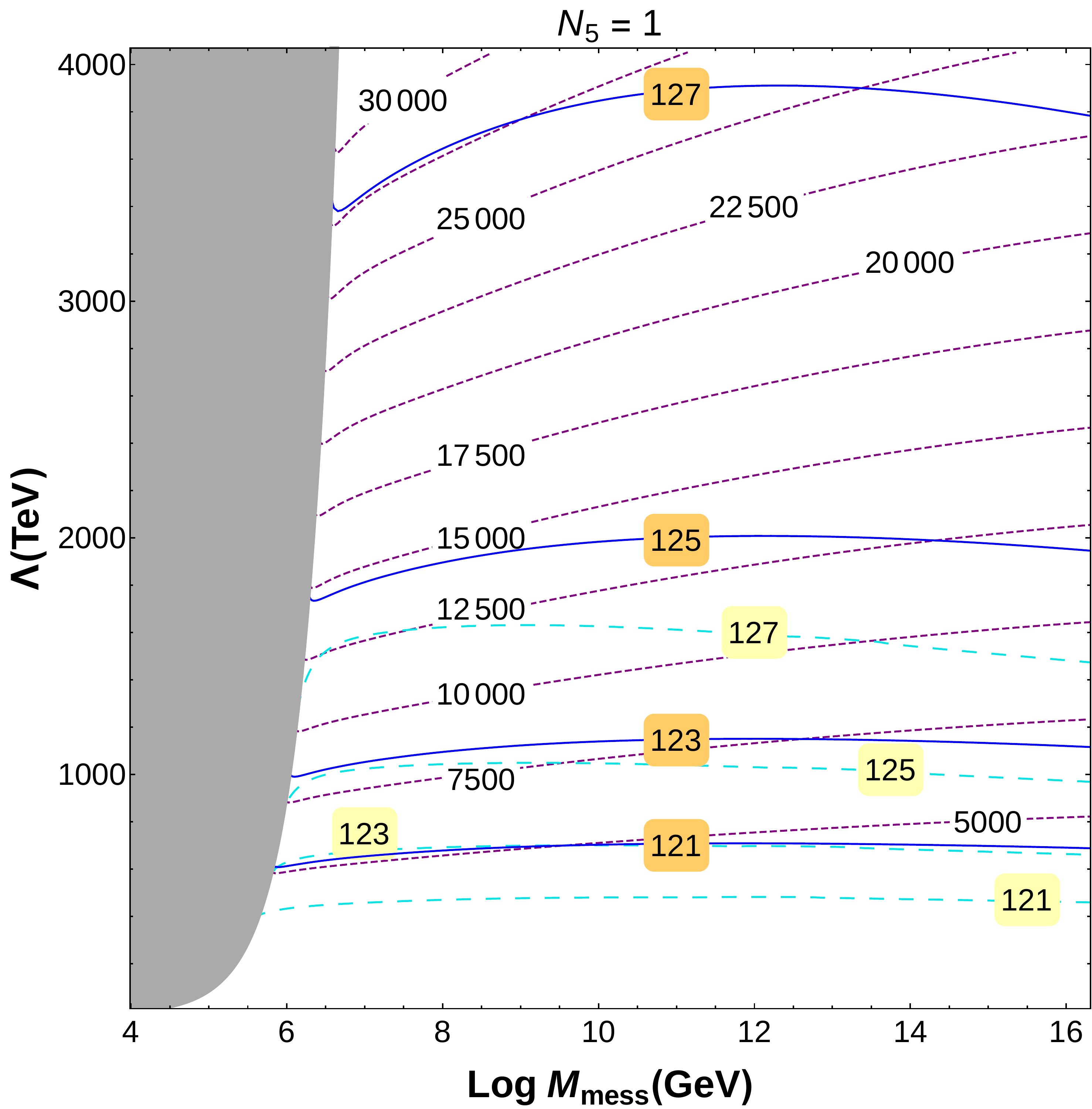}
\includegraphics[width=.49\linewidth]{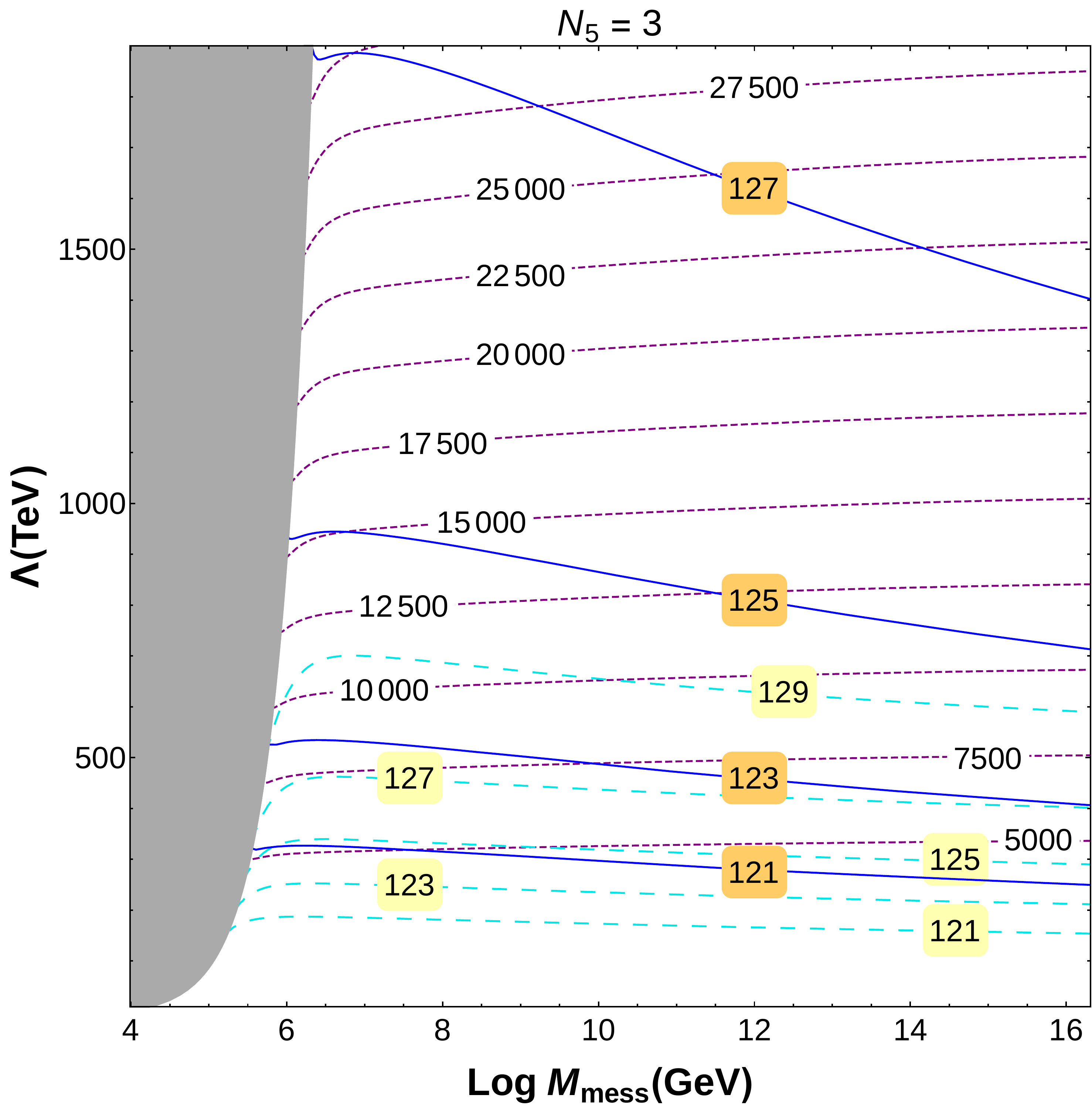}
\caption{Contours of the average stop mass (purple lines) and the Higgs mass in the $\Mm$--$\Lambda$ plane for the minimal GMSB and different choices of $N_5$. 
The dashed cyan lines and the solid blue lines correspond to the Higgs mass calculated with \texttt{FeynHiggs} and \texttt{SusyHD}, respectively.
The $\Lambda \geq \Mm$ region is shaded in grey.}
\label{fig:minGMSB1}
\end{figure*}

However, there is a drawback that make this solution unworkable. As is clear from the right panel of figure~\ref{fig:minGMSB1}, the value of $\Lambda$ required to produce heavy enough stops, so that the Higgs mass is consistent with the experimental one, is rather large, $\Lambda\simeq {\cal O}(10^6)$ GeV. Consequently, the focus-point solution occurs for $\Lambda > \Mm$, which leads to negative mass-squared for some (scalar components of) messengers, that is not acceptable (it would lead to charge and color breaking). 

\begin{figure*}[ht]
\centering 
\includegraphics[width=.49\linewidth]{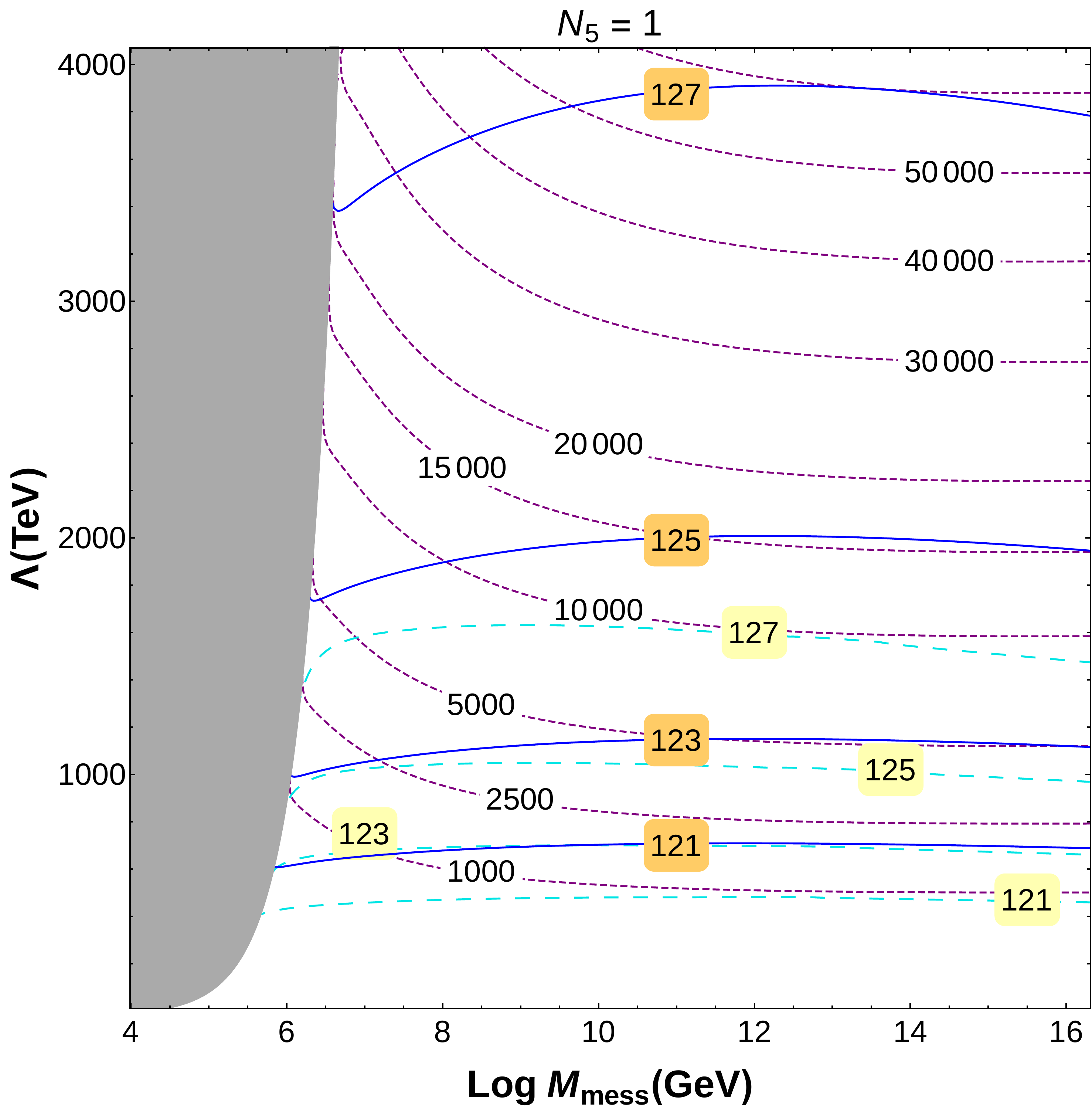}
\includegraphics[width=.49\linewidth]{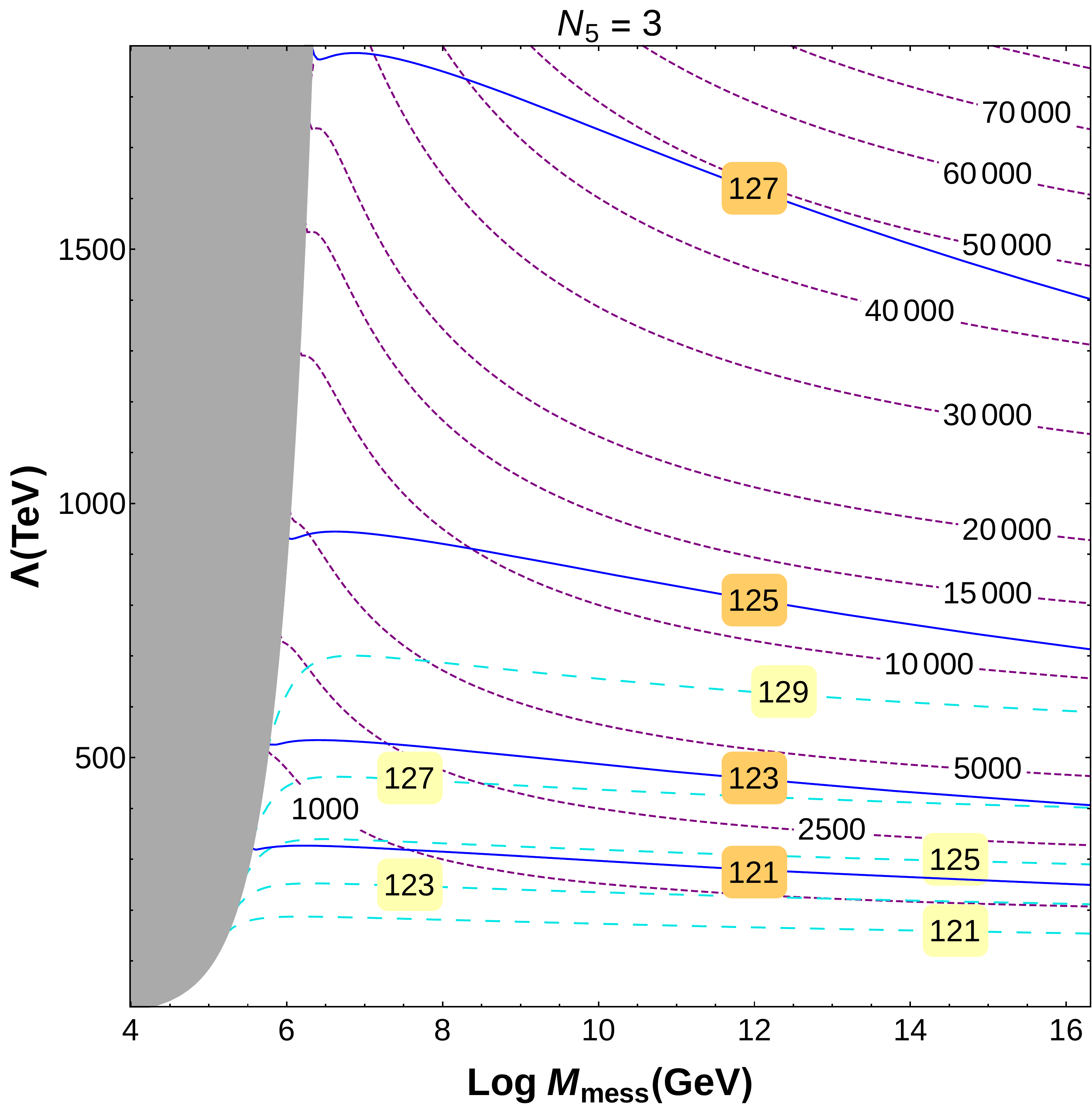}
\caption{Contour lines of constant $\Delta_{\Lambda}$ (purple lines) and the Higgs mass calculated with \texttt{FeynHiggs} (dashed cyan lines) and \texttt{SusyHD} (solid blue lines) in the $\Mm$--$\Lambda$ plane, for minimal GMSB and different values of $N_5$. 
The unphysical region, $\Lambda \geq \Mm$, is shaded in grey.}
\label{fig:minGMSB2}
\end{figure*}

\begin{figure*}[ht]
\centering 
\includegraphics[width=.49\linewidth]{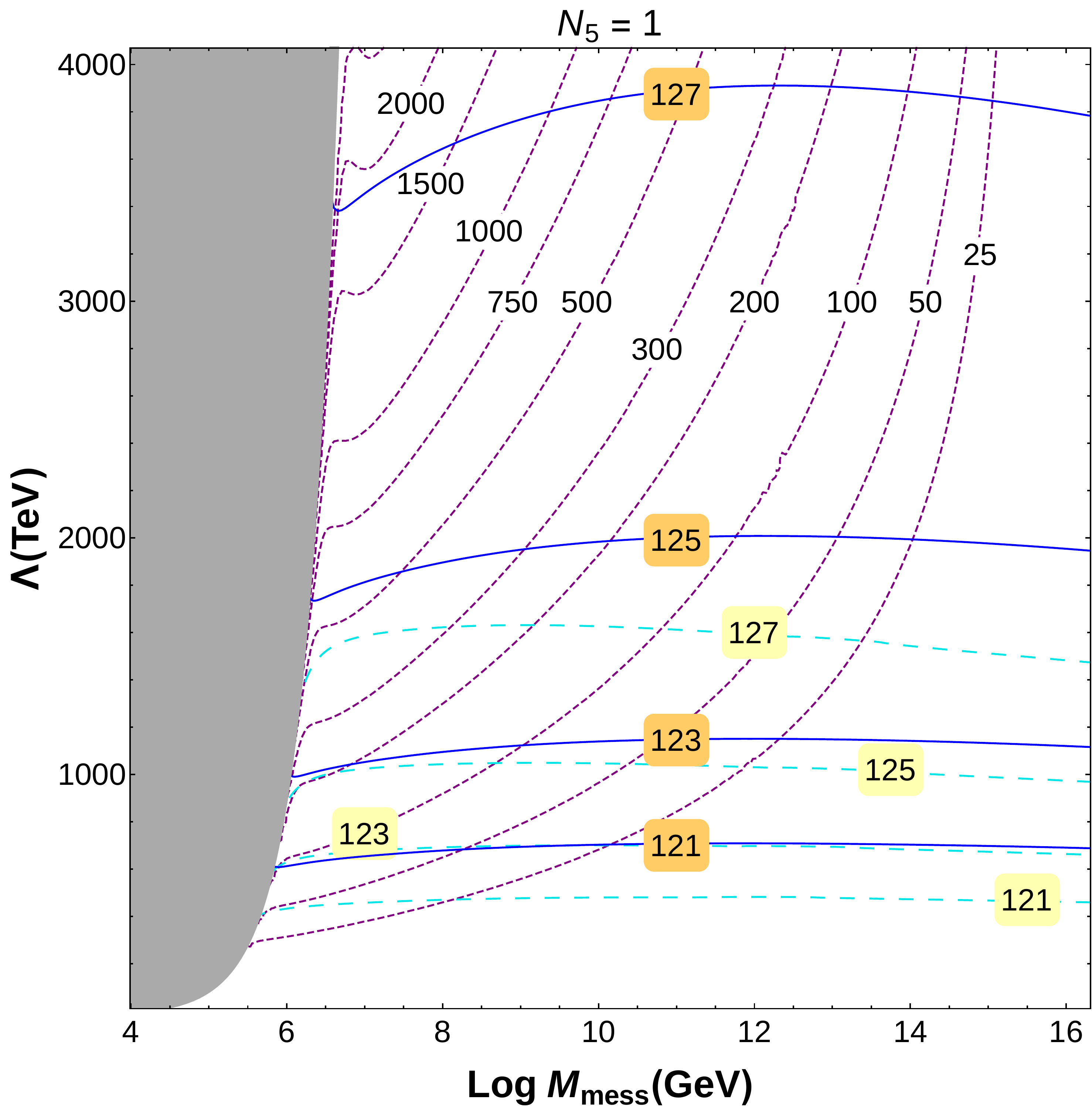}
\includegraphics[width=.49\linewidth]{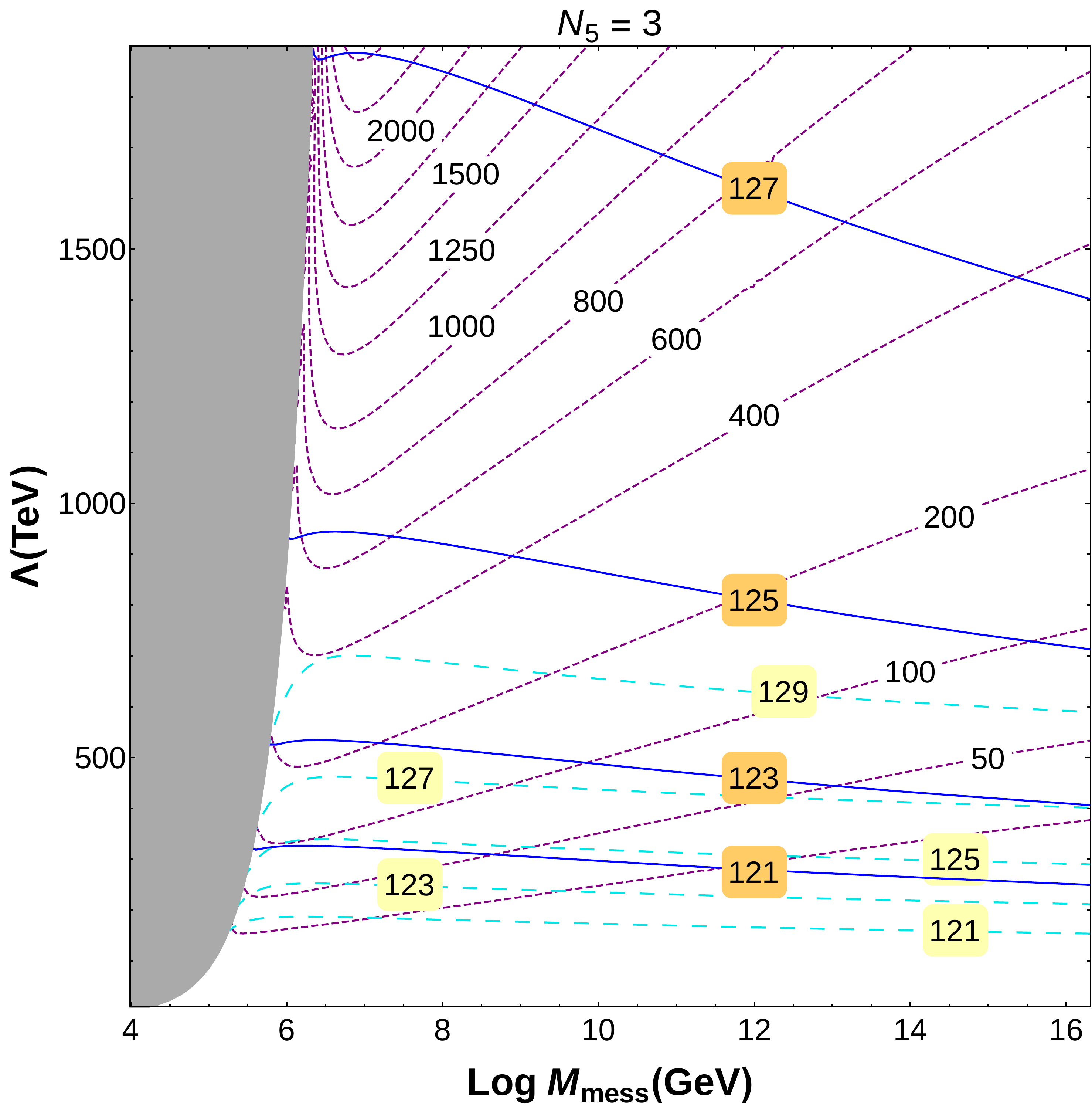}
\caption{Contours of $\Delta_{\Mm}$ (purple lines) and the Higgs mass computed with \texttt{FeynHiggs} (dashed cyan lines) and \texttt{SusyHD} (solid blue lines) in the $\Mm$--$\Lambda$ plane, for minimal GMSB and different choices of $N_5$.
The grey shaded region corresponds to $\Lambda \geq \Mm$.}
\label{fig:minGMSB3}
\end{figure*}

Figures~\ref{fig:minGMSB1}, \ref{fig:minGMSB2} and \ref{fig:minGMSB3} show  $\overline{m}_{\tilde{t}}$, $\Delta_{\Lambda}$ and $\Delta_{\Mm}$ respectively, in the acceptable region (which does not include the focus-point for the above-mentioned reason), 
for two choices of the number of messengers, namely $N_5=1$ and $N_5=3$, and $\tan\beta=10$. The figures have been obtained using 
the complete expressions for the initial values of the soft masses given in \ref{MGMSBappendix}. The two fine-tunings were evaluated along the lines of subsection~\ref{tunGMSB}, with the RG-parameters computed as in  ref.~\cite{Casas:2014eca}. 
\footnote{The only difference is 
the LE scale that was now chosen to be $10$ TeV.}

Notice that the fine-tuning associated to $\Mm$, which is an independent parameter in this context, is always lower than $\Delta_{\Lambda}$.

The bottom line is that the electroweak fine-tuning for the minimal GMSB is very large, ${\cal O}(10^4)$ for $m_h=125$ GeV, evaluated with \texttt{SusyHD}, although it can be below $10^3$ if we compute $m_h$ with \texttt{FeynHiggs} and allow $m_h=123$ GeV, to account for theoretical uncertainties. From now on we will take this conservative value for $m_h$ to compare the fine-tuning of different scenarios, independently of the code used to compute $m_h$.
 The main cause of the large fine-tuning is the small value of $A_t$, which leads to large stop masses in order to reproduce the experimental Higgs mass. Those stop masses require in turn a large value of $\Lambda$, increasing the value of $m_{H_u}^2({\rm LE})$ and thus the fine-tuning. The latter is actually more severe than previous estimates in the literature since previous codes used to evaluate the Higgs mass did not work with enough accuracy for large stop masses.

Consequently, in order to make the GMSB scenario less fine-tuned one has to go beyond this minimal setup, exploring mechanisms to incorporate a non-vanishing $A_t$. This is the subject of the next two sections.

\section{Models with radiatively generated $\bf{A}$-terms\label{sec:4}}

The possibility of generating $A$\hyp{}terms through loops thanks to messenger\hyp{}MSSM interactions has been analyzed in many papers; see refs.
\cite{Chacko:2001km, Chacko:2002et, Shadmi:2011hs,
Evans:2011bea, Jelinski:2011xe, Evans:2012hg, Craig:2012xp,Albaid:2012qk, Abdullah:2012tq, Perez:2012mj, Endo:2012rd,
Evans:2013kxa,Calibbi:2013mka}. In ref.~\cite{Evans:2013kxa}, Evans and Shih performed an extensive survey  of this type of models. Namely, they considered both, scenarios with cubic MSSM\hyp{}MSSM\hyp{}messenger or MSSM\hyp{}messenger\hyp{}messenger operators in the superpotential. Besides, they distinguished between cases where the relevant messengers are squark\hyp{}like or Higgs\hyp{}like. They concluded that all scenarios had fine\hyp{}tunings to the sub\hyp{}percent level. Actually, all scenarios analyzed had tunings at the sub\hyp{}permil level, except one, based on the coupling 
\be
\label{Evans}
\Delta W = \lambda\ UH_u\phi_{10, Q},
\ee
which had $\Delta\simeq 850$. Here, $\phi_{10,Q}$ denotes the $Q$-like component of a messenger in the 10 representation of $SU(5)$ (for further details see ref.~\cite{Evans:2013kxa}). The corresponding one-loop-generated $A$-term can be read from \ref{CPZappendix}. The mentioned fine-tuning represents an appreciable improvement over the minimal GMSB model, analyzed in the previous section, where the minimal fine-tuning was slightly above 1000, and typically was $\gsim 2500$. 

Nevertheless, in order to compare the performance of both models, the tuning must be evaluated with the same criteria. Here, we re-analyzed the Evans and Shih model defined by eq.~(\ref{Evans}) in an improved fashion, consistent with the analysis of the minimal case (section~\ref{sec:3}). First of all, 
we include the exact two-loop corrections to the scalar masses, whereas in ref.~\cite{Evans:2013kxa} these corrections were approximated by the first term in their $x$-expansion ($x$ has been defined in eq.~(\ref{x})); see \ref{CPZappendix} for further details. Second, as discussed in previous sections, in order to calculate the Higgs mass we have used the more recent versions of \texttt{FeynHiggs} and \texttt{SusyHD}, while the authors of ref.~\cite{Evans:2013kxa} used the \texttt{SOFTSUSY} code  \cite{Allanach:2001kg}. It is known that, for given supersymmetric parameters, \texttt{SOFTSUSY} produces a larger value for $m_h$ especially when the stops are heavy \cite{Bagnaschi:2014rsa,Vega:2015fna}, which is the typical case in GMSB. Consequently, their results are more optimistic than ours. On the other hand, we have allowed the theoretical value of $m_h$ to be as low as $123$ GeV to account for theoretical uncertainties, whereas in ref.~\cite{Evans:2013kxa} $m_h$ was fixed at 125 GeV. Finally, the fine-tuning criterion in ref.~\cite{Evans:2013kxa} was also (slightly) different. Instead of considering $\Lambda$ as an independent parameter, and thus evaluating $\Delta_\Lambda$, they considered a bunch of independent parameters: $\Lambda_i=\{g_i^2\Lambda^2, y_i^2\Lambda, \lambda \Lambda, \Lambda_{1-loop}\}$ (see ref.~\cite{Evans:2013kxa} for the precise definition of $\Lambda_{1-loop}$). Certainly, the various contributions to $m^2_{H_u}({\rm LE)}$ are proportional to the various (squared) couplings in the theory (gauge couplings, Yukawa couplings and the $\lambda$-coupling) times $\Lambda$. In this sense,  their criterion captures the level of ``conspiracy'' between different terms. However, all those couplings (except $\lambda$) are fixed by experiments, and it is contrived to examine variations of parameters which are fixed \cite{Ciafaloni:1996zh, Casas:2005ev, Cheung:2012qy}. Then, the possible variations of all those terms arise from those of $\Lambda$ and are thus correlated. This criterion normally increases the fine-tuning, since a certain variation in $\log\Lambda$ modifies more the value of $\log m^2_{H_u}({\rm LE})$ than the same variation in $\log\Lambda_i$.

\begin{figure}[ht]
\centering 
\includegraphics[width=1.\linewidth]{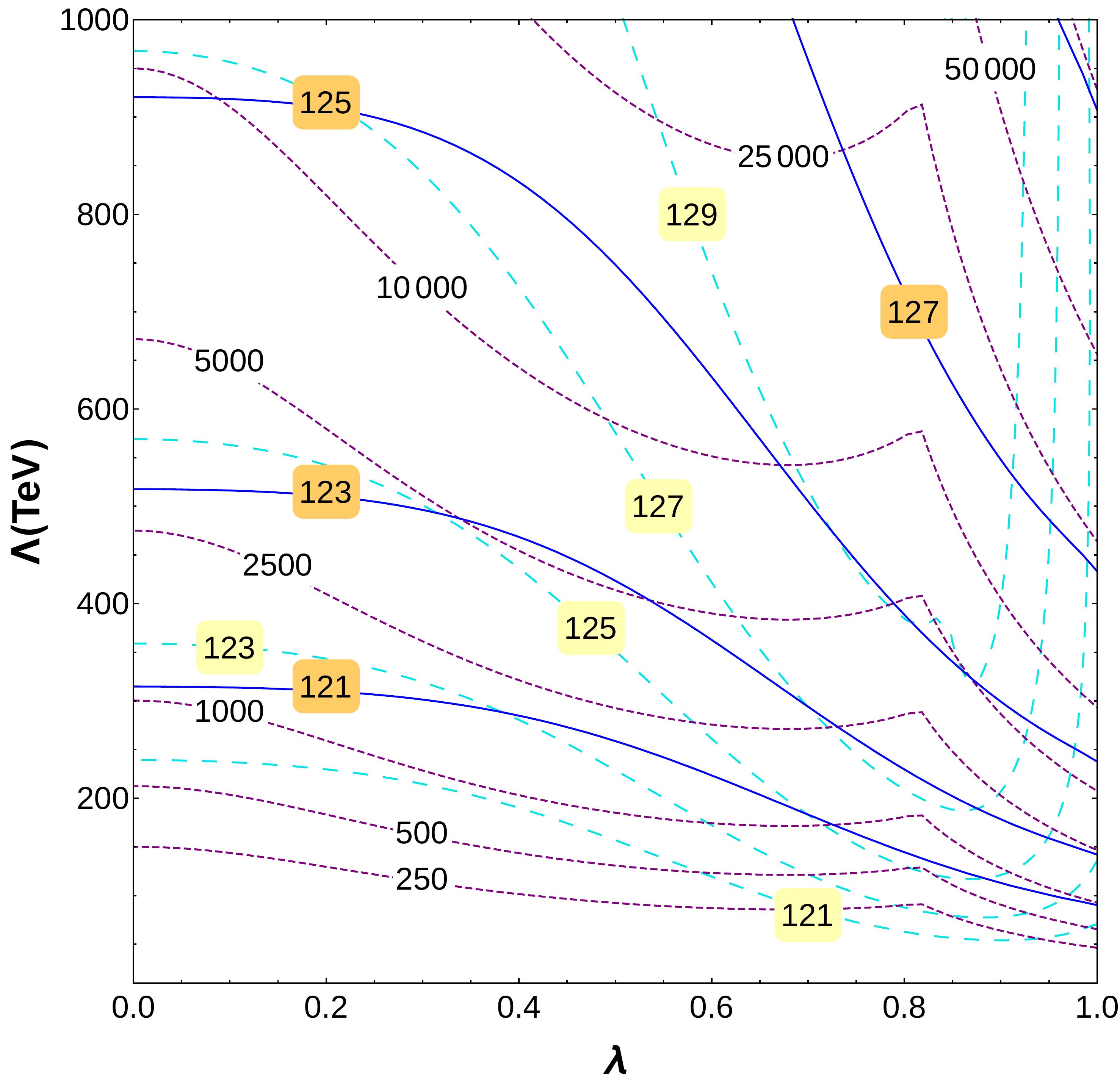}
\caption{Contours of $\Delta$, defined as Max$\{\Delta_i\}$(purple lines) and the Higgs mass computed with \texttt{FeynHiggs} (dashed cyan lines) and \texttt{SusyHD} (solid blue lines) in the $\lambda$--$\Lambda$ plane, for the model defined in eq.~(\ref{Evans}), $\Mm=10^8$ GeV and $N_{10}=1$.}
\label{fig:Evansplot}
\end{figure}

Fig.~\ref{fig:Evansplot} shows the fine-tuning, $\Delta$ in the plane $\lambda-\Lambda$, evaluated according to our criterion for the model defined by the extra term (\ref{Evans}) and $M_{\rm HE}=M_{\rm mess}=10^8$ GeV, $N_{10}=1$ and $\tan\beta=10$. For small $\lambda$,  the fine-tuning is dominated by $\Delta_\Lambda$, while for large $\lambda$ by $\Delta_\lambda$; thus the kink in the contour lines. The minimal fine-tuning is about 1500 when $m_h$ is computed with \texttt{SusyHD}, and close to 250 when computed with \texttt{FeynHiggs}. This represents a certain improvement w.r.t. the minimal GMSB.

In a later paper, Calibbi et al.\ (CPZ)~\cite{Calibbi:2013mka} considered a version of radiatively generated $A$-terms. More precisely,  they considered the following term in the messenger superpotential:
\begin{eqnarray}
\label{extraW3}
\Delta W =  \lambda_U Q_3U_3 \Phi_{H_u} \ ,
\end{eqnarray}
where $Q_3, U_3$ are the third generation of quark superfields and $\Phi_{H_u}$ is the SU(2) doublet (included in the messenger superfield $\Phi$) with the same quantum numbers as $H_u$. Again, a trilinear scalar coupling for the stops is generated at one loop, see \ref{CPZappendix}.
One can imagine that the $\lambda_U$ coupling is related in some way to the standard Yukawa couplings, which justifies neglecting similar terms for other superfields. 
In addition to the trilinear coupling, there appear new contributions to the scalar masses \cite{Evans:2011bea,Calibbi:2013mka} at one loop and two loops; see \ref{CPZappendix}. 

Next we re-visit this model in greater detail, to show the obstacles to reduce the fine-tuning in this kind of scenarios.
For large enough $\lambda_U$ (not far from the top Yukawa coupling), the generated $A_t$-term can have the appropriate size at low energy to maximize the threshold correction to the Higgs mass or, in other words, to minimize the magnitude of the stop masses in order to reconcile $m_h$ with the experimental value. 
According to CPZ, the requirement $m_h>123$ GeV can be fulfilled for much lighter stops than in the minimal GMSB model. They find that, for $N_5=1$, the approximate optimal choice is $\lambda_U\simeq 0.7$, for which the lightest stop can be as light as 400 GeV if the messenger mass is suitably chosen. 
The second stop, however, is much heavier, close to 2 TeV. Consequently, the $\Lambda$ scale might be much lower than in the minimal GMSB, which apparently would amount to a substantial reduction in the electroweak fine-tuning. In addition, the rest of the super-particles (squarks, gluinos, etc.) are also closer to the LHC reach since their masses are proportional to $\Lambda$. It should be mentioned here that CPZ used \texttt{SOFTSUSY} to compute $m_h$, so, as discussed above, their conclusions are in the optimistic range.

Nevertheless, this scenario has some shortcomings. Due to the new contributions to scalar masses, $m_{H_u}^2$ gets a negative correction, which can be very important. As a consequence the low-energy (absolute) value of $m_{H_u}^2$ tends to be larger, which implies a more severe electroweak fine-tuning in eq.~(\ref{minh}). Hence, the reduction of the fine-tuning due to the lighter stops (and thus smaller $\Lambda$) is compensated by this effect. Actually, CPZ noted that, in spite of having lighter stops, the value of $\mu$, and thus the fine-tuning, does not decrease appreciably.
Besides, compared with the minimal GMSB scenario, the model contains an extra parameter, namely the $\lambda_U$ coupling. Since we do not know the theoretical connection of this with the other parameters of the model, $\{\Lambda, M_{\rm mess}, \mu, B\}$, its fine-tuning parameter, $\Delta_{\lambda_U}$, should be computed and considered in eq.~(\ref{BG}), as we actually did for the Evans and Shih model above. As we will see soon, for large values of $\lambda_U$, which CPZ consider interesting for LHC phenomenology, the value of  $\Delta_{\lambda_U}$ becomes very important and even larger than $\Delta_{\Lambda}$.

Concerning the stop masses, according to CPZ, the choice $\lambda_U=0.75$ leads to stops as light as possible. We have checked that for that value of $\lambda_U$, stops are lighter than in the minimal GMSB scenario, though the effect is not dramatic. Demanding $m_h>123$ GeV requires stops above $\sim 6$~TeV (if $m_h$ is computed with \texttt{SusyHD} or $\sim4$~TeV with \texttt{FeynHiggs}), somewhat smaller than for the minimal GMSB. 

\begin{figure*}[ht]
\centering 
\includegraphics[width=.49\linewidth]{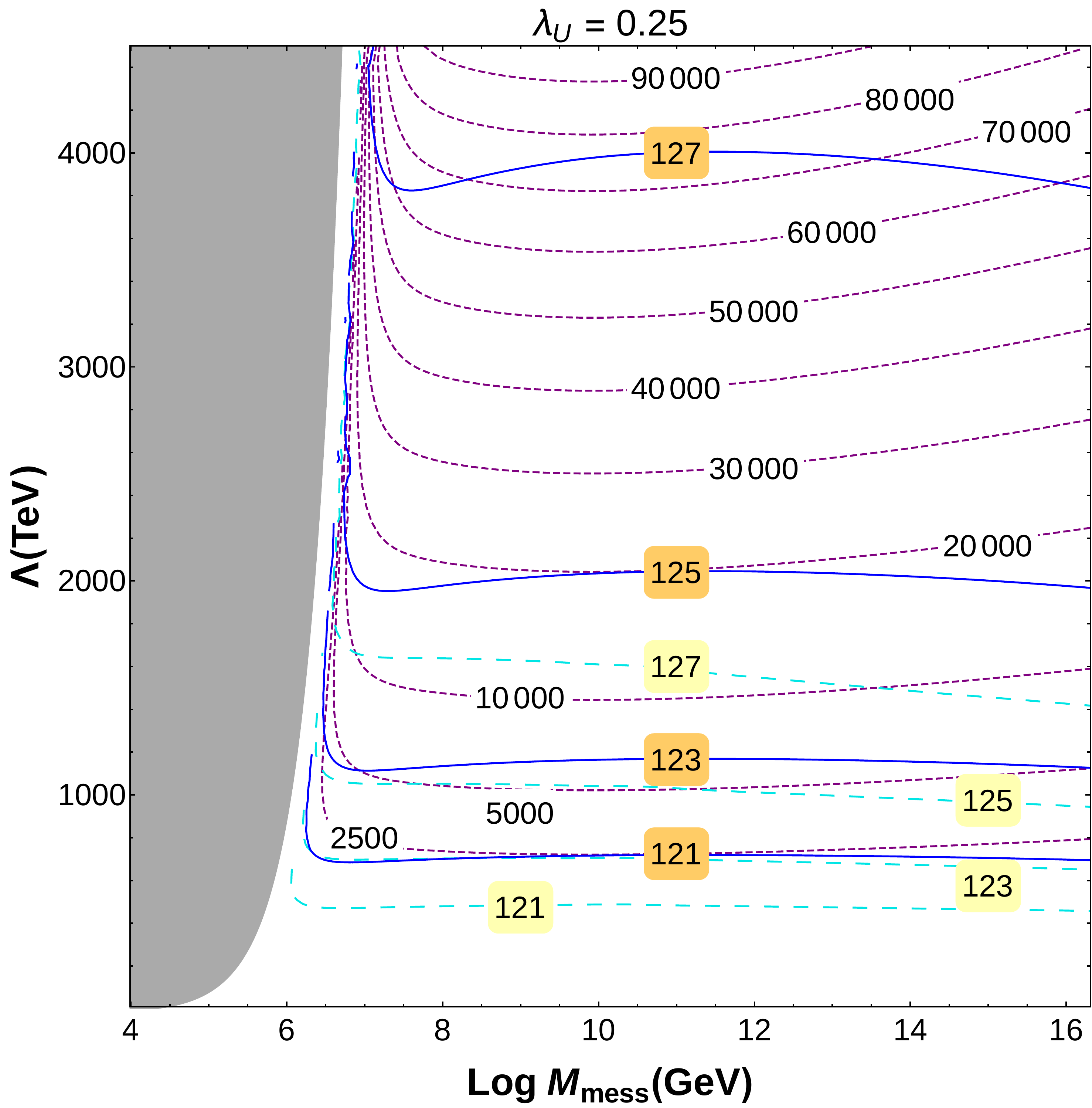}
\includegraphics[width=.49\linewidth]{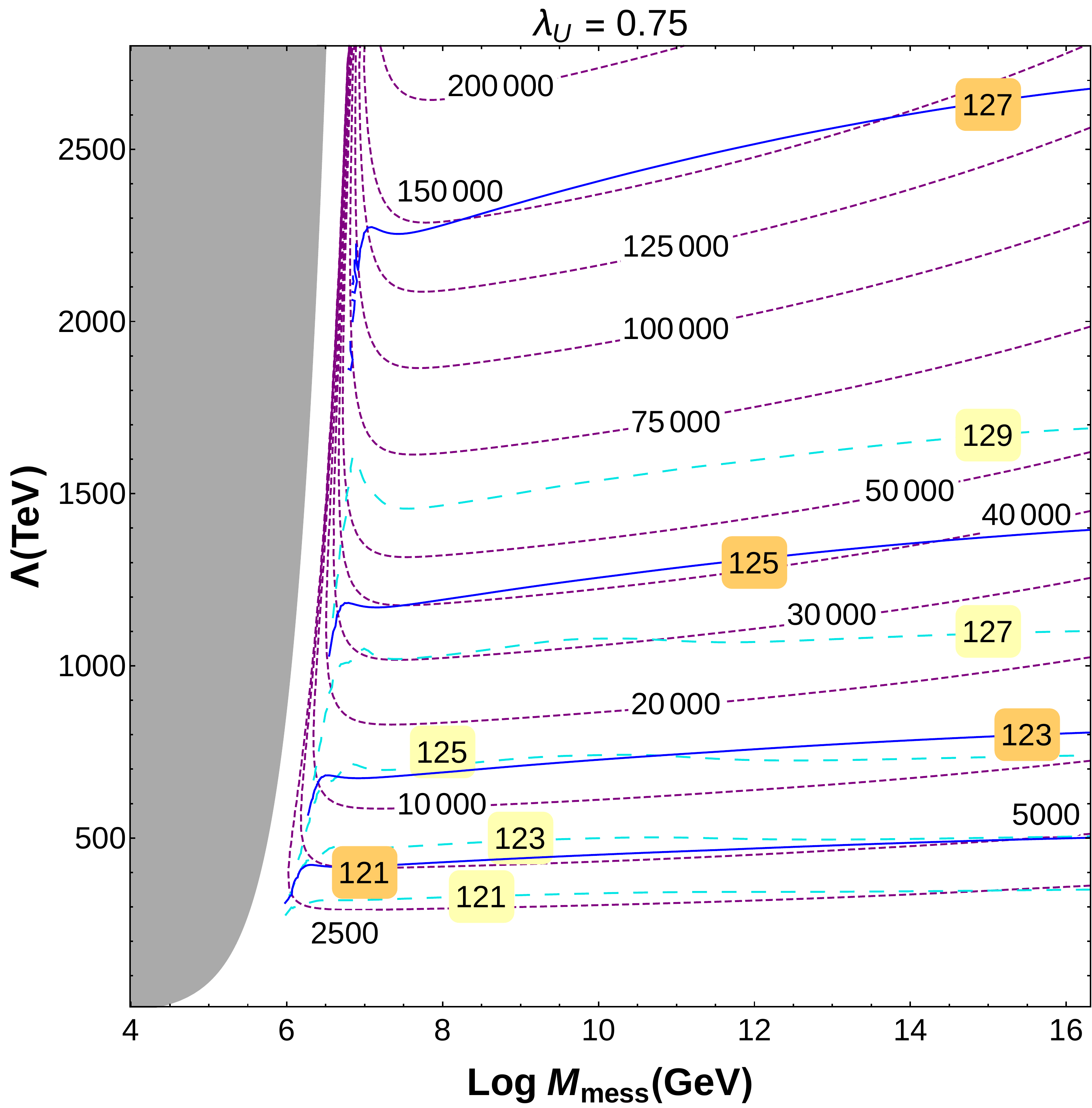}
\caption{Contours of constant $\Delta_{\Lambda}$ and the Higgs mass calculated with \texttt{FeynHiggs} (dashed cyan lines) and \texttt{SusyHD} (solid blue lines) in the $\Mm$--$\Lambda$ plane, for the CPZ model and two different choices of $\lambda_U$. 
The grey shaded region corresponds to $\Lambda \geq \Mm$.}
\label{fig:Calibbi2}
\end{figure*}

\begin{figure*}[ht]
\centering 
\includegraphics[width=.49\linewidth]{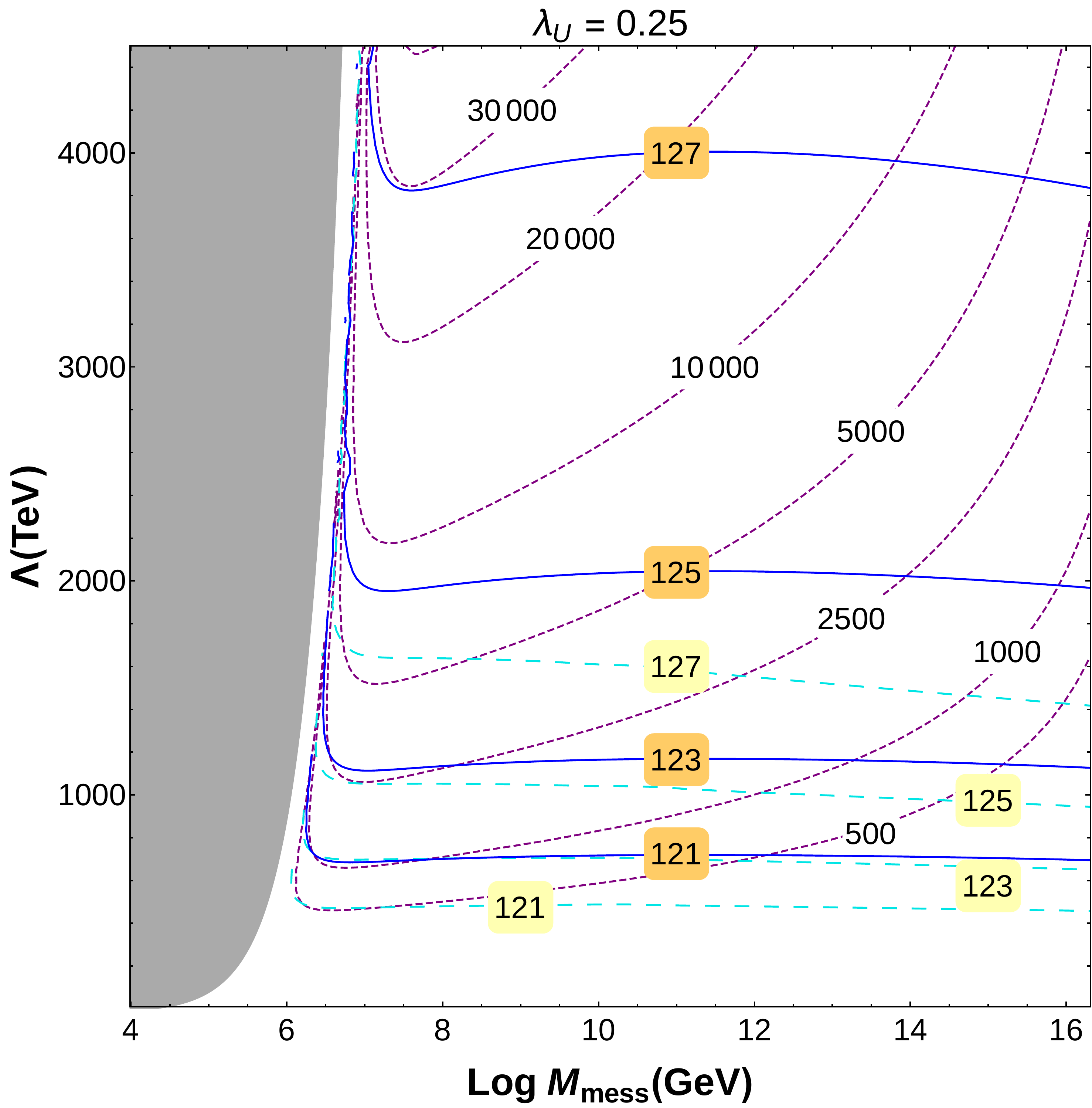}
\includegraphics[width=.49\linewidth]{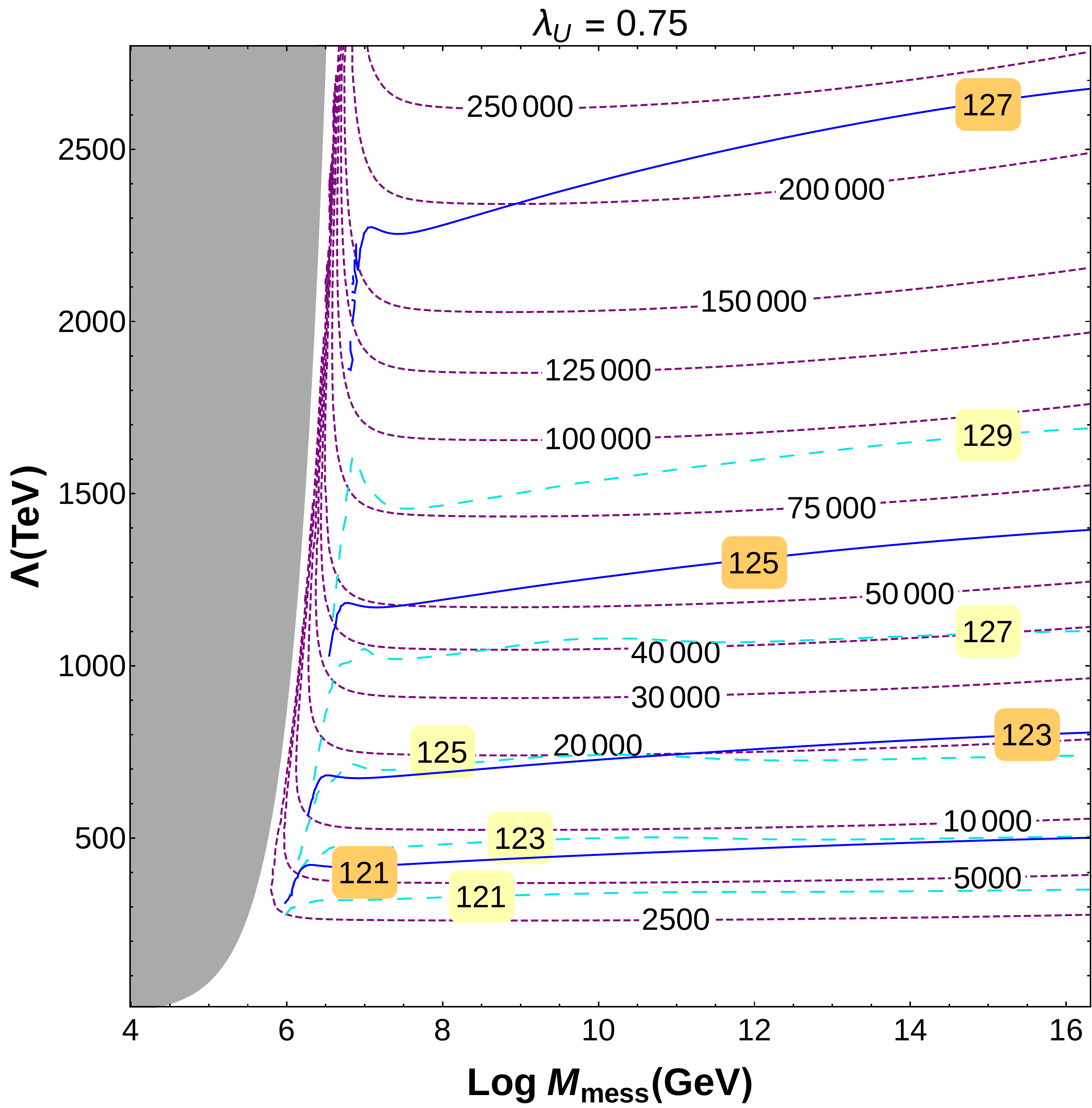}
\caption{Contour lines of $\Delta_{\lambda_U}$ and the Higgs mass computed with \texttt{FeynHiggs} (dashed cyan lines) and \texttt{SusyHD} (solid blue lines) in the $\Mm$--$\Lambda$ plane for the CPZ model and two different values of $\lambda_U$. 
The unphysical region, $\Lambda \geq \Mm$, is shaded in grey.}
\label{fig:Calibbi3}
\end{figure*}

Figures~\ref{fig:Calibbi2} and \ref{fig:Calibbi3} illustrate some of the previous points. To avoid proliferation of plots, we have focused on the $N_5=1$ case, but the results for other values of $N_5$ are analogous. In addition, we have fixed $\tan\beta=10$ and $M_{{\rm LE}}=10$ TeV.
Note that the figures
show a ``threshold line'' close to the $\Lambda\geq\Mm$ (grey) region, which cannot be crossed. This virtual line signals when a stop mass-squared gets negative. Notice here that when $\Lambda$ approaches $\Mm$ the initial values of the soft terms get important contributions, which are negligible otherwise; see \ref{CPZappendix}. 

Concerning fine-tuning things get worse. Figure~\ref{fig:Calibbi2} shows the fine-tuning associated to $\Lambda$ in the $\Mm$--$\Lambda$ plane for $\lambda_U=0.25, 0.75$. While for $\lambda_U=0.25$ the fine-tuning ($\sim 6000$ with \texttt{SusyHD}, $\sim 2500$ with \texttt{FeynHiggs}) is only slightly worse than in the minimal GMSB, for $\lambda_U= 0.75$ it becomes more than two times worse. So it does not pay off to go to large values of $\lambda_U$, even if the stops become lighter.

Actually, for large $\lambda_U$, the fine\hyp{}tuning associated to $\lambda_U$ itself becomes even bigger than that associated to $\Lambda$. This can be checked in figure~\ref{fig:Calibbi3}. While for $\lambda_U=0.25$ the $\Delta_{\lambda_U}$\hyp{}parameter is large, but smaller than $\Delta_\Lambda$, and thus can be ignored; for $\lambda_U=0.75$ it becomes larger. The situation becomes worse as $\lambda_U$ is increased.

\begin{figure*}[ht]
\centering 
\includegraphics[width=.49\linewidth]{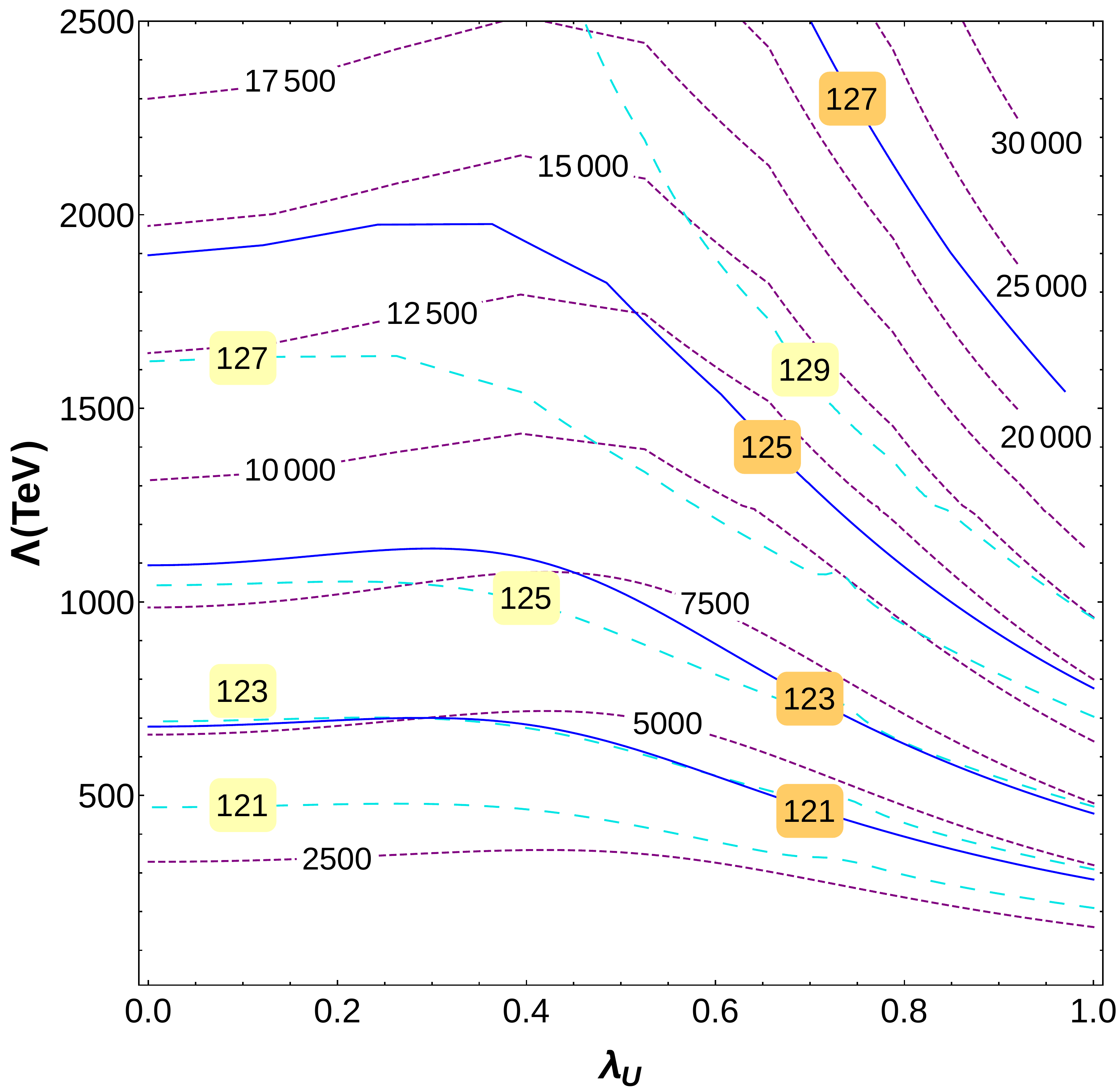}
\includegraphics[width=.49\linewidth]{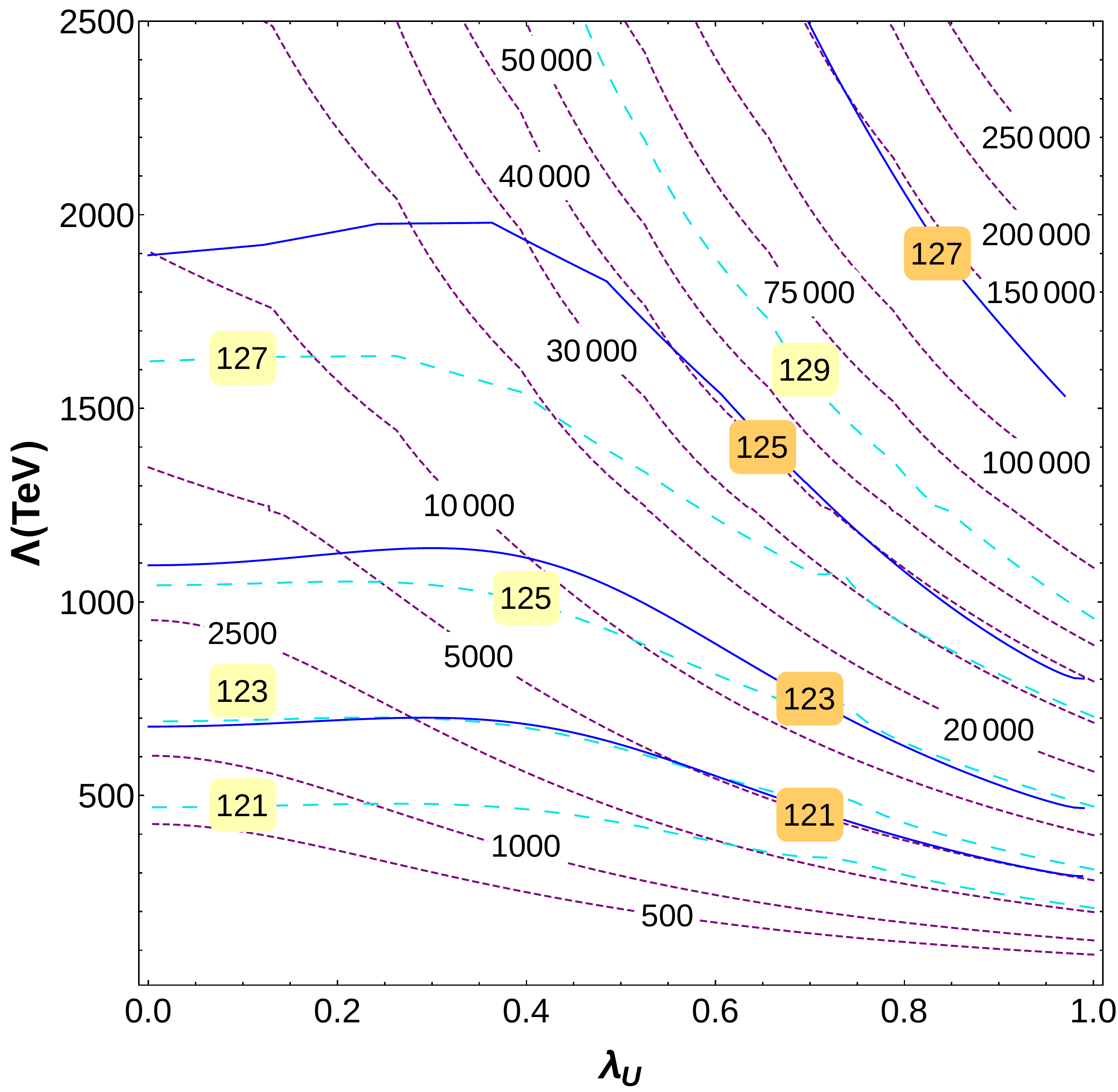}
\caption{Left: contours of the constant average stop mass and the Higgs mass calculated with \texttt{FeynHiggs} (dashed cyan lines) and \texttt{SusyHD} (solid blue lines) in the $\Lambda$--$\lambda_U$ plane for the CPZ model, $\Mm=10^8$ GeV and $N_{5}=1$. 
Right: contour lines of $\Delta_{\Lambda}$ and Higgs mass in the same plane.}
\label{fig:Calibbi4}
\end{figure*}

The previous discussion is well summarized by figure~\ref{fig:Calibbi4}. In the left panel of figure~\ref{fig:Calibbi4}, we show contour lines of constant $m_h$ and (averaged) $m_{\tilde t}$ in the $\lambda_U$--$\Lambda$ plane for a fixed value of the messenger mass, namely $\Mm=10^8$ GeV (for other choices of $\Mm$ the results are essentially equivalent). It can be noted that the stop mass is minimized for $\lambda_U\sim 0.7$, in agreement with CPZ results. 
The right panel of figure~\ref{fig:Calibbi4} shows contour lines of constant fine-tuning, i.e.\ the $\Delta_\Lambda$ parameter. Clearly, the best choice is $\lambda_U=0$, i.e.\ the minimal GMSB scenario. As a matter of fact, for $\lambda_U\simgt 0.55$ 
the fine-tuning associated to $\lambda$ becomes dominant, i.e. $\Delta_{\lambda_U}>\Delta_\Lambda$, so the situation gets even worse.
On the other hand, notice also that for large $\lambda_U$ lines are cut. This is due to the (left-handed) slepton masses falling below the present bounds. This imposes an absolute bound on the size of $\lambda_U$.

The final conclusion is that, generically, $A$-terms generated radiatively thanks to the couplings of messengers to the observable fields in the superpotential fail to improve appreciably the fine-tuning of the minimal GMSB model. In some cases they lead to a milder fine-tuning but hardly better than the  one per mil level.

\section{A simple scenario\label{sec:5}}
 
In this section, we consider a simple GMSB scenario that in principle can get a fine-tuning as mild as possible. The model is a variation of the idea put forward by Basirnia et al.\ in ref.~\cite{Basirnia:2015vga}, namely, the generation of the desired sizeable $A_t$-term by the exchange of messengers at tree level.

Let us start with the usual GMSB superpotential, eq.~(\ref{Wmess}), enhanced with two additional terms,
\begin{equation}
\label{extraW4}
\Delta W = (k X + \Mm) \Phi_{H_u}\Phi_{H_d}+ \lambda Q_3U_3 \Phi_{H_u} + \lambda' X H_u \Phi_{H_d} \ .
\end{equation}

The term proportional to $\lambda$ coincides with the one used in the CPZ model, but now there is an extra term, proportional to $\lambda'$, which directly couples the $X$ superfield to the messengers and the standard Higgs fields. As mentioned in section~\ref{sec:1}, one can always re-define the scalar component of $X$ so that $\langle X\rangle=0$. Now, however, such a re-definition would induce extra terms in $W$. So, we will assume for simplicity that the $X$ superfield, which couples to $H_u$ and $\Phi_{H_d}$ as in eq.~(\ref{extraW4}) with no additional terms in $W$, has a small VEV compared to $\Mm$. Then, one can eliminate the heavy messengers using $\frac{\partial W }{\partial \Phi_{H_u}} =\frac{\partial W }{\partial \Phi_{H_d}}=0 $. The resulting effective superpotential reads
\begin{eqnarray}
\label{extraW5}
\Delta W_{\mathrm{eff}} \sim -\frac{\lambda\lambda'}{k X+ \Mm} X  Q_3 H_u U_3 \ .
\end{eqnarray}
Expanding in powers of $X$ and replacing it by its scalar component, $\langle X\rangle$, we get a small correction to the standard top Yukawa term, $Y_t \rightarrow Y_t -\lambda\lambda' \frac{\langle X\rangle}{\Mm}$. Replacing $X$ by its  $F$-component, we get a trilinear scalar coupling for the stops, with a coefficient
\begin{eqnarray}
\label{extraAtnew}
y_tA_t=  -{\lambda\lambda'}\frac{F_X}{\Mm} \left[1 + {\cal O}({\langle X\rangle}/{\Mm})\right]\simeq -\frac{\lambda\lambda'}{k}\Lambda \ .
\end{eqnarray}

Note that the generated $A_t$-term arises at tree level, so the combination of couplings $\lambda \lambda'/k$ must be small. 

The above model is a modification of the model proposed in ref.~\cite{Basirnia:2015vga}. The main difference is that the authors of ref.~\cite{Basirnia:2015vga} got a sufficiently small $A_t$ by assuming that there was a second spurion, $X'$, with $F_{X'}<F_X$, which was the field coupled as in eq.~(\ref{extraW4}) (see their eq.~(1.3)). Here we show that, in fact, this is not necessary. One can live just with one spurion, provided the $\lambda'$ coupling (which was implicitly assumed to be $\lambda'=1$ in ref.~\cite{Basirnia:2015vga}) is small enough. This represents a conceptual simplification.

Unfortunately, as stressed in ref.~\cite{Basirnia:2015vga}, after integrating out the $\Phi_{H_u},\Phi_{H_d}$ superfields, one not only obtains the modified superpotential eq.~(\ref{extraW5}), but also a modified K\"ahler potential, $K$. Namely, replacing $\Phi_{H_u}=-\lambda'XH_u/M$ in the canonical $K$, one gets a term
\begin{eqnarray}
\label{Keff}
\Delta K_{\rm eff} =\frac{|\lambda'|^2}{M^2} |X|^2|H_u|^2 \ .
\end{eqnarray}
in the effective K\"ahler potential, which leads to an extra contribution to $m_{H_u}^2$,
\begin{eqnarray}
\label{deltamHu}
\delta m_{H_u}^2 =-\left|\frac{\lambda'F_X}{M}\right|^2 =-\left|\frac{\lambda'\Lambda}{k}\right|^2\ .
\end{eqnarray}
Comparing eq.~(\ref{extraAtnew}) with eq.~(\ref{deltamHu}), we see that it is not possible to arrange the parameters so that $\delta m_{H_u}^2$ is small, since
\begin{eqnarray}
\label{rate}
\left|\frac{(y_tA_t)^2}{\delta m_{H_u}^2}\right| = |\lambda|^2\ .
\end{eqnarray}

\begin{figure*}[ht]
\centering 
\includegraphics[width=.49\linewidth]{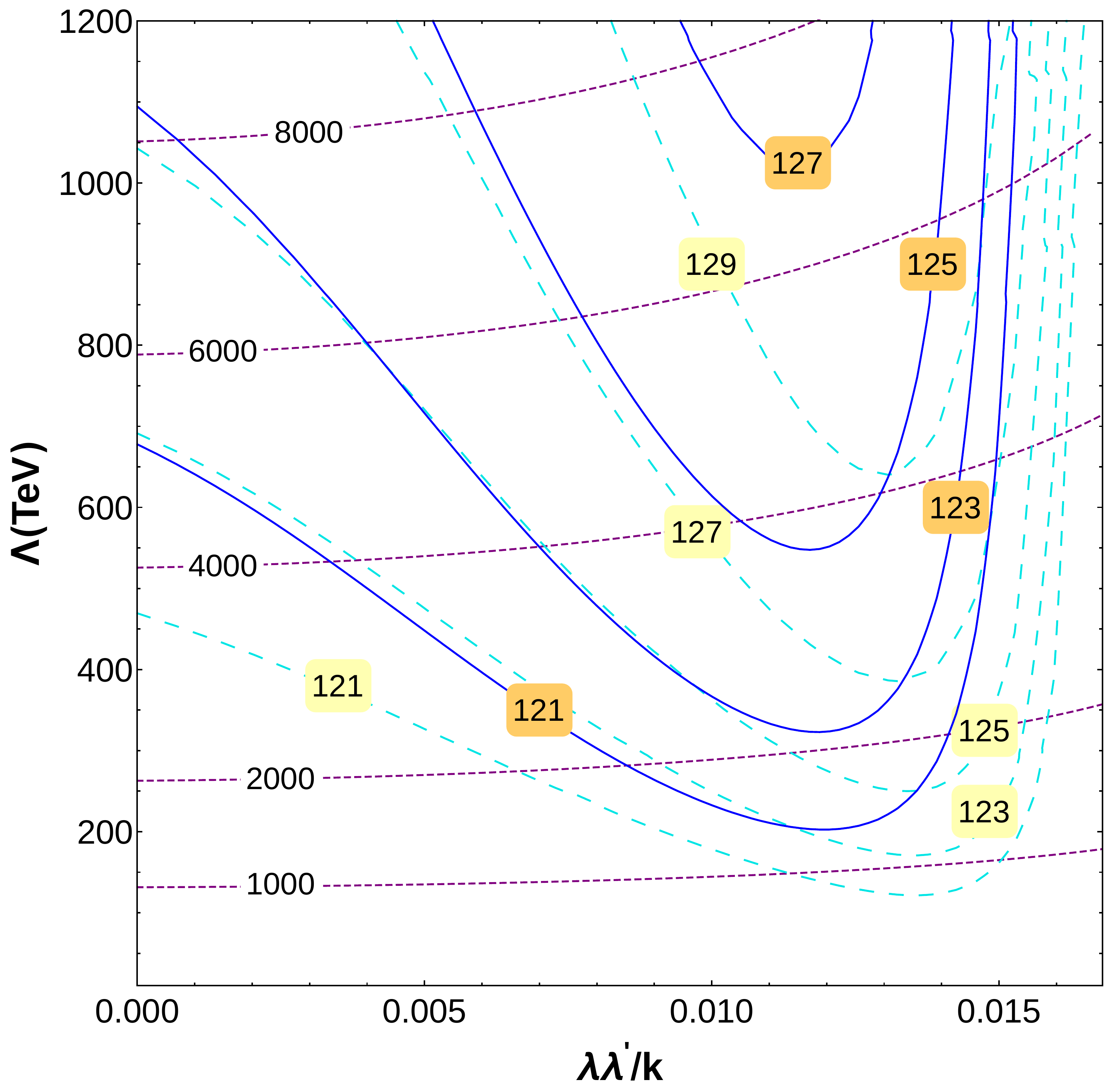}
\includegraphics[width=.49\linewidth]{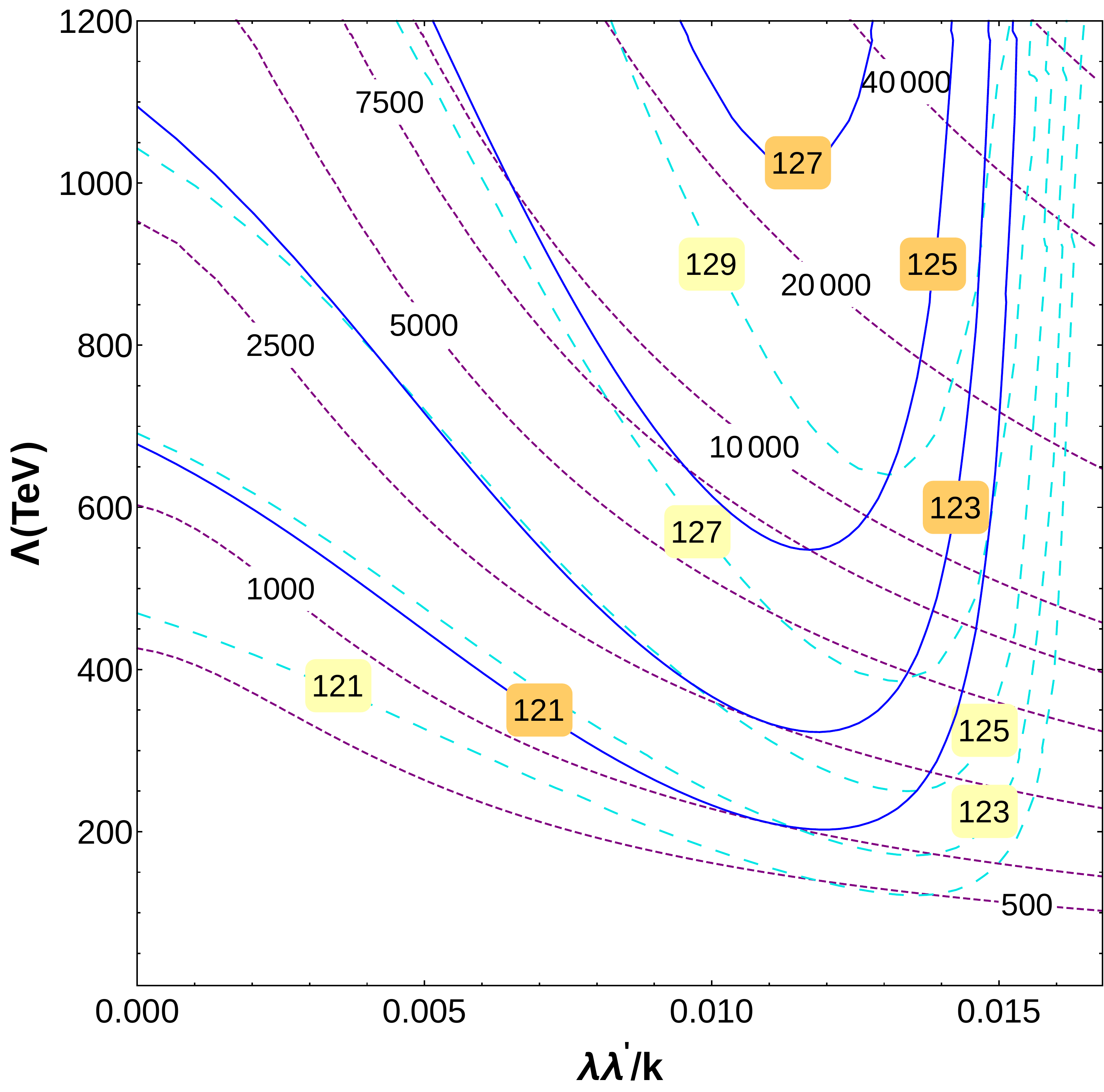}
\caption{Model of eq.~(\ref{extraW4}). Left panel: contour lines of the average stop mass and the Higgs mass calculated with \texttt{FeynHiggs} (dashed cyan lines) and \texttt{SusyHD} (solid blue lines) in the $\Lambda$--$\frac{\lambda\lambda'}{k}$ plane for $\Mm=10^8$ GeV, $\lambda=1.5$ and $N_{5}=1$. 
Right panel: contours of $\Delta_{\Lambda}$ and Higgs mass in the same plane.}
\label{fig:Ourproposal}
\end{figure*}

In other words, a sizeable $y_t A_t$ implies a sizeable and negative $\delta m_{H_u}^2$. Such a result is a manifestation of the so-called ``little $A_t^2/m^2$ problem'' discussed in \cite{Craig:2012xp}, i.e.\ the fact that a large $A$-term is normally accompanied by a similar or larger sfermion mass-squared.
This is bad news for naturalness since the fine-tuning in $\Lambda$ is proportional to $|m_{H_u}|^2$, see eq.~(\ref{BGLambda}). Consequently, for a given value of $y_t A_t$ one should minimize the (negative) size of $\delta m_{H_u}^2$ as much as possible. One obvious way is to consider a large $\lambda$, without spoiling the perturbativity regime, $\lambda \leq {\cal O}(1)$. In contrast, increasing the number of messengers does not help since there is always a unique combination of them that couples to $Q_3 U_3$ in eq.~(\ref{extraW4}).
This scenario is illustrated in figure~\ref{fig:Ourproposal} for  $M_{\rm mess}=10^8$ GeV, $N_{5}=1$ and $\lambda=1.5$. One can see that, for a given value of $m_h$, there is a value of the $\lambda \lambda'/k$  combination that nearly minimizes the stop masses and the fine-tuning. 
Assuming, as usual, a $\sim 2$ GeV theoretical uncertainty in the determination of $m_h$, it turns out that the average stop mass can drop to 2.2~TeV while the fine-tuning can be $\sim 2500$. Comparing with the minimal GMSB for the same messenger mass (figures~\ref{fig:minGMSB2} and \ref{fig:minGMSB3}), we see that stop masses can be much smaller, though the fine-tuning does not appreciably improve. In fact, it is clearly worse than for the Evans and Shih model of eq.~(\ref{Evans}), see Fig.~\ref{fig:Evansplot}, although it is 
much better than for the CPZ model. 

An alternative, and improved situation occurs when the relevant messengers are not $\Phi_{H_u},\Phi_{H_d}$, but $\Phi_{u}, \Phi_{\bar u}$, with the same quantum numbers as $U_3, \overline U_3$ \cite{Basirnia:2015vga}. In this case, the relevant superpotential is similar to that of eq.~(\ref{extraW4}),
\begin{eqnarray}
\label{extraW6}
\Delta W = (k X + \Mm) \Phi_{u}\Phi_{\bar u}+ \lambda Q_3 \Phi_{u}H_u + \lambda' X U_3 \Phi_{\bar u} \ ,
\end{eqnarray}
and the effective superpotential, after integration of $\Phi_{u}, \Phi_{\bar u}$, reads exactly as that of eq.~(\ref{extraW5}). Now, the effective K\"ahler potential leads to a negative contribution to the singlet squark mass-squared, $\delta m_{\tilde u_3}^2$ (instead of $\delta m_{H_u}^2$),
with the same size as before, i.e.\ the r.h.s.\ of eq.~(\ref{deltamHu}). This is much less dangerous than the previous $\delta m_{H_u}^2$. Actually, reducing the size of the stop masses lowers the final absolute value of $m_{H_u}^2$, thus alleviating the fine-tuning. 

However, as above, the size of this negative contribution to $\delta m_{\tilde u_3}^2$ is limited by the requirement of perturbativity for $\lambda$. Actually, we have also to ensure that the VEVs of the coloured scalar fields are vanishing. This implies in particular that the final value of $m_{\tilde u}^2$ should be kept positive, which entails an upper bound on $A_t$, namely
\begin{eqnarray}
\label{limitstab1}
|\lambda' F| < m_{\tilde u_3}^{(0)}\Mm\hspace{0.5cm}\Rightarrow \hspace{0.5cm}|y_t A_t|<\lambda m_{\tilde u_3}^{(0)} \ ,
\end{eqnarray}
where $m^2_{\tilde u_3}=(m_{\tilde u_3}^{(0)})^2 + \delta m_{\tilde u_3}^2$, i.e. $m_{\tilde u_3}^{(0)}$ is the standard value of the minimal version of GMSB.
We can refine this analysis by studying the square-mass matrix for the $\{\phi_{\bar u}, \phi_{u}, \tilde u_3\}$ fields, i.e. before integrating out the messengers. This can be obtained from the scalar potential associated to the superpotential (\ref{extraW6}):
\bea
M^2_{u}=
\left(
\begin{array}{ccc}
\Mm^2\ \ \ \ \ \ \ & (kF_X)^\dagger & \\
(kF_X)\ \ \ \ \ \ \  & \Mm^2 & (\lambda' F_X)^\dagger  \\
 & (\lambda' F_X)& (m_{\tilde u}^{(0)})^2 \\
\end{array} \right) \ .
\eea
Stability requires the two order-two minors of the above matrix to be positive. This coincides with the stability conditions eqs.~(\ref{xlimit}, \ref{x}) and (\ref{limitstab1}), respectively. In addition, we must demand the determinant to be positive,
\begin{eqnarray}
\label{limitstab2}
\Mm^2\left[ \Mm^2 (m_{\tilde u}^{(0)})^2 - |\lambda' F|^2\right] - |kF|^2 (m_{\tilde u}^{(0)})^2 \ >\  0
\end{eqnarray}
This condition becomes relevant if any of the other two conditions are nearly saturated.

Another aspect of this setup is that the messengers cannot belong anymore to the $(5+\bar 5)$, since this does not accommodate an $U_3$ field. Now, we have to consider copies of $(10 + \overline{10})$, although again only one combination of messengers contributes to the above $\lambda Q_3 \Phi_{u}H_u$ coupling.

The fact that the $(10 + \overline{10})$ representation contains pieces with the same quantum numbers as $Q_3, \overline{Q}_3$, say $\Phi_{Q}, \Phi_{\bar Q}$, allows for additional possibilities. In particular, one can add new pieces, $\lambda \Phi_{Q} U_3H_u + \lambda' X Q_3 \Phi_{\bar Q}$,
to the superpotential in eq.~(\ref{extraW6}) (for simplicity, we assume that the couplings have the same size as above). Then, after integrating out all the messengers, both $m_{\tilde U_3}^2$, $m_{\tilde Q_3}^2$ receive the same negative contribution given by the r.h.s.\ of eq.~(\ref{deltamHu}). However, now $y_tA_t$ is two times larger than before, so expression (\ref{rate}) becomes
\begin{eqnarray}
\label{rate2}
\left|\frac{(y_tA_t)^2}{\delta m_{\tilde t}^2}\right| = 4 |\lambda|^2\ ,
\end{eqnarray}
thus improving substantially the $(y_tA_t)^2/\delta m_{\tilde t}^2$ ratio. 
A final possibility, which actually optimizes the $(y_tA_t)^2/\delta m^2$ ratio, is to consider messengers in the $(5 + \overline{5})+(10 + \overline{10})$. Playing just with a messenger in each representation does not conflict with perturbativity of the gauge couplings for any value of $M_{\rm mess}$. Then, one can use three messengers in the $\Phi_{H_u}+ \Phi_{H_d}$, $\Phi_{u}+ \Phi_{\bar u}$ and $\Phi_{Q}+ \Phi_{\bar Q}$ representations, coupled as above to generate effective contributions to $y_t A_t$.
Assuming again that the $\lambda$, $\lambda'$ couplings are the same for all of them, we get a ``universal'' shift in $\delta m_{H_u}^2=\delta m_{\tilde U_3}^2= \delta m_{\tilde Q_3}^2$, given by the r.h.s.\ of eq.~(\ref{deltamHu}), while the generated  $y_tA_t$ term is three times bigger, so
\begin{eqnarray}
\label{rate3}
\left|\frac{(y_tA_t)^2}{\delta m^2}\right| = 9 |\lambda|^2\ .
\end{eqnarray}

\begin{figure*}[ht]
\centering 
\includegraphics[width=.49\linewidth]{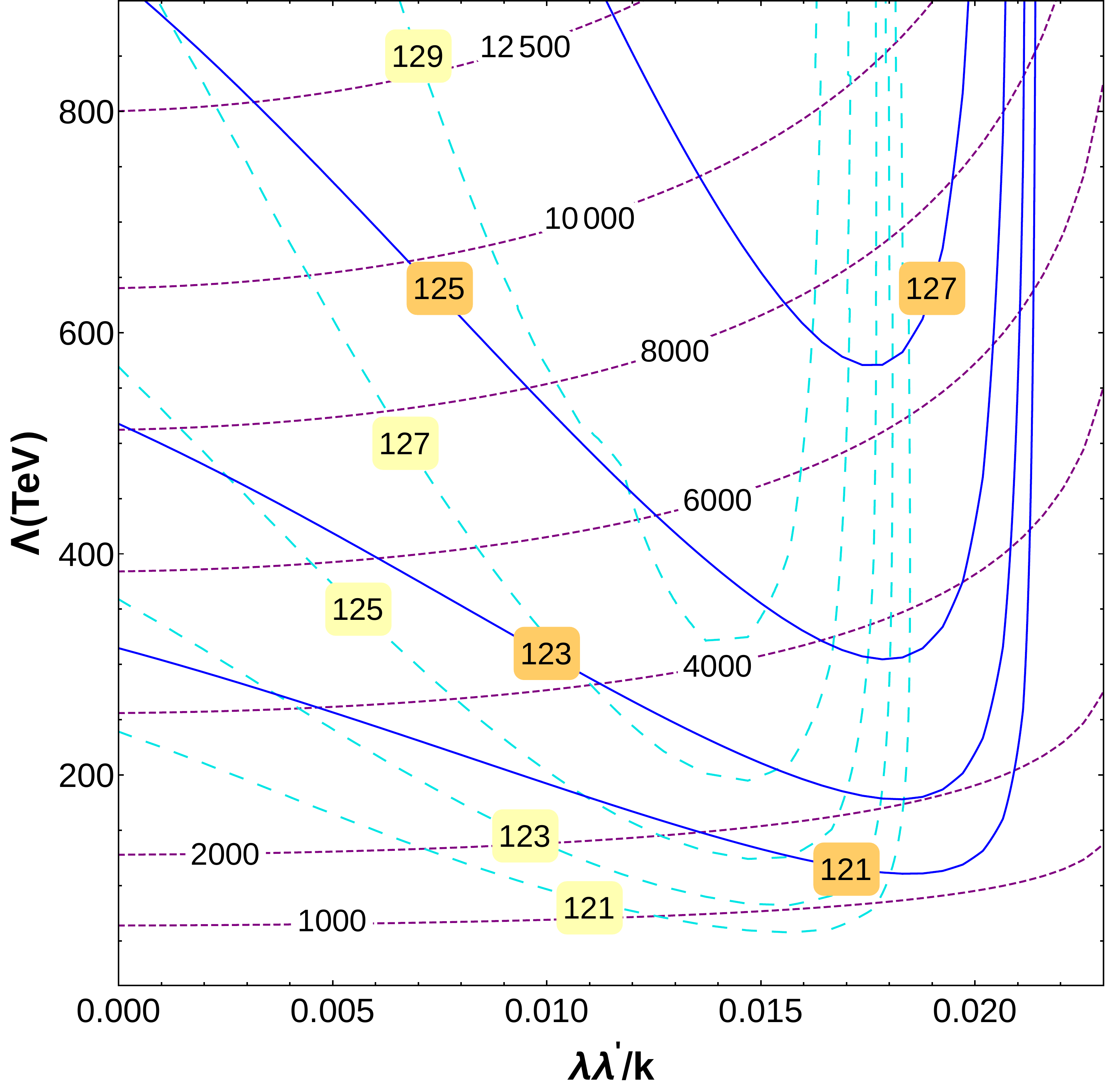}
\includegraphics[width=.49\linewidth]{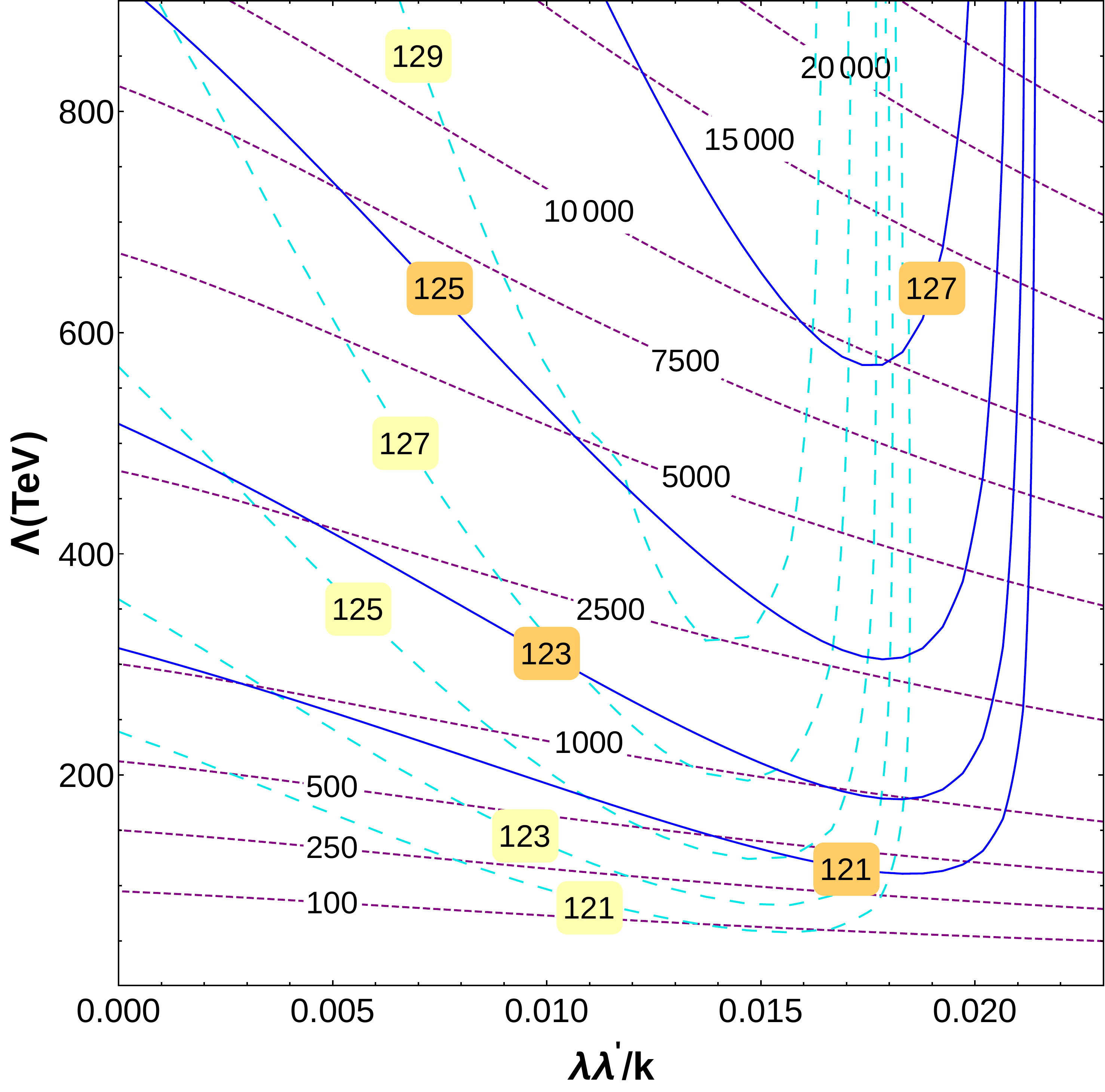}
\caption{The same as figure~\ref{fig:Ourproposal}, but for the model of eq.~(\ref{extraW7}).}
\label{fig:NonMFV}
\end{figure*}

Nevertheless, improving the $(y_tA_t)^2/\delta m_{\tilde t}^2$ ratio does not necessarily leads to a milder fine-tuning. As mentioned above, a negative shift in the initial value of  $m_{H_u}^2$ increases the fine-tuning, while the same shift in $m_{\tilde u_3}^2$ or $m_{\tilde Q_3}^2$ reduces it, since these mass-squared terms  enter  in the RG shift of $m_{H_u}^2$ with negative sign. Thus, the scenarios corresponding to eqs.~\eqref{rate2} and \eqref{rate3} do not improve the fine-tuning with respect to the scenarios with just one messenger $\Phi_{u}+ \Phi_{\bar u}$ coupled as in eq.~(\ref{extraW6}). For the same reason, the scenario that actually optimizes the fine-tuning is the one where the messenger that induces the $A$-term is $\Phi_{Q}+ \Phi_{\bar Q}$, rather than $\Phi_{u}+ \Phi_{\bar u}$, because  $m_{\tilde Q_3}^2$ enters the RG shift of $m_{H_u}^2$ with a larger (negative) coefficient than $m_{\tilde u_3}^2$ \cite{Casas:2014eca}. Thus, in this optimized model the superpotential reads
\begin{eqnarray}
\label{extraW7}
\Delta W = (k X + \Mm) \Phi_{Q}\Phi_{\bar Q}+ \lambda \Phi_{Q} U_3 H_u + \lambda' X Q_3 \Phi_{\bar Q} \ ,
\end{eqnarray}
and  the generated $y_t A_t$, $\delta m_{\tilde Q_3}^2$ pieces are given by the r.h.s.\ of eqs.~\eqref{extraAtnew} and \eqref{deltamHu}. 
This scenario is illustrated in figure~\ref{fig:NonMFV}, for $M_{\rm mess}=10^8$ GeV, $N_{10}=1$ and $\lambda=1.5$. The pattern is similar to the setup of eq.~(\ref{extraW4}), illustrated in figure~\ref{fig:Ourproposal}, but now one can get substantially milder fine-tuning. Namely, evaluating the Higgs mass with \texttt{SusyHD} (\texttt{FeynHiggs}), for $m_h>123$ GeV the fine-tuning may drop below $1000$ (250). This is even better (though not dramatically) than the optimum model with radiatively generated $A$-terms identified by Evans and Shih, i.e. that of eq.~(\ref{Evans}), illustrated in Fig.~\ref{fig:Evansplot}; and  hence it is an optimal GMSB model concerning fine-tuning (and playing with minimal matter content).

Let us finally mention that in the kind of scenarios discussed in this section, there is an additional fine-tuning source associated to the new parameters, in particular to the combination $\lambda \lambda'/k$, which is proportional to the initial value of $y_t A_t$. We have checked  that this contribution to the fine-tuning is smaller than that associated with $\Lambda$.

\section{Conclusions\label{sec:6}}
 
Models with gauge-mediated supersymmetry breaking 
have become one of the most popular supersymmetric scenarios, especially for their prevention of dangerous FCNC effects. However, these models typically present a high degree of fine-tuning, due to the initial absence of top trilinear scalar couplings, $A_t=0$. This makes the threshold correction to the Higgs mass, $m_h$, to be far from its maximal value, so the stop masses must be quite large in order to generate sizeable radiative corrections to $m_h$, able to reconcile its value with the experimental one. Such large stop masses (around 10 TeV) imply in turn a large value of $\Lambda \sim F/\Mm$ and thus a severe fine-tuning in order to get the right electroweak breaking scale.

In this paper, we have carefully evaluated the fine\hyp{}tuning associated with GMSB, using also the most recent codes for the computation of the Higgs mass in the MSSM, which plays a relevant role for such an evaluation. We show that previous analyses underestimated the fine\hyp{}tuning of GMSB. The actual one is typically of the order a few per ten thousand in the minimal model. Then, we have examined some proposals which have been made in the literature to improve the situation, incorporating a mechanism to generate the $A_t$ term, while keeping the minimal observable matter content.
They always involve non-trivial couplings between the messengers and the MSSM superfields in the superpotential.

We find that, even though the stops can be made lighter, this does not necessarily lead to a better fine-tuning. In particular, in the model proposed by Calibbi et al.~\cite{Calibbi:2013mka}, an $A_t$\hyp{}coupling is generated at one loop, so that the stops can indeed be lighter than in the minimal version of GMSB. Nevertheless, we show that the fine-tuning gets actually worse, essentially due to the additional contributions to the scalar masses (especially $m_{H_u}^2$). Things are better, however, for the model with radiatively generated $A_t$, proposed by Evans and Shih, which was the most favourable one from an extensive survey of models of this kind \cite{Evans:2013kxa}.

On the other hand, in the scenario proposed by 
 Basirnia et al.\ in ref.~\cite{Basirnia:2015vga}, the $A_t$-term is generated at tree level and the prospects are generically better. We explore this scenario, proposing a modified (and conceptually simplified) version which is arguably the optimum in the GMSB setup (with minimal matter content) concerning the fine-tuning issue. In this model, the fine-tuning can be better than one per mil. This is still a severe fine-tuning, but substantially milder than in other versions of GMSB, and of the same order as in other MSSM constructions. 
 
We also explore the so-called ``little $A_t^2/m^2$ problem'' \cite{Craig:2012xp}, i.e.\ the fact that a large $A_t$-term is normally accompanied by a similar or larger sfermion mass-squared, which typically implies an increase in the fine-tuning. We find the version of GMSB for which this ratio is as large as possible, namely ${\cal O}(10)$. However, we show that the model that optimizes this ratio does not coincide with the one that has the smallest fine-tuning.
 
\vspace{0.2cm}

\begin{acknowledgements}
We thank B. Zaldivar for useful discussions and collaboration during the first stage of this work. We also thank S. Heinemeyer for providing us the last version of \texttt{FeynHiggs} and his assistance in its usage.
This work has been partially supported by the MICINN, Spain, under contract FPA2013-44773-P, Consolider-Ingenio CPAN CSD2007-00042, as well as MULTIDARK CSD2009-00064. 
We also thank the Spanish MINECO {\em Centro de excelencia Severo Ochoa Program} under
Grant SEV-2012-0249. 
S.R.  is supported by the Campus of Excellence UAM+CSIC and  K.R. by the Spanish Research Council (CSIC) within the JAE-Doc program.
\end{acknowledgements}

\appendix
\section{GMSB high-energy spectrum}

\subsection{Minimal GMSB}
\label{MGMSBappendix}
In the minimal GMSB, gauginos acquire their mass at one loop. Computing the corresponding Feynman diagrams, the gaugino masses are \cite{Dimopoulos:1996gy,Martin:1996zb,Poppitz:1996xw}:
\be
M_i = \df{\alpha_i}{4\pi}\Lambda N_5 g(x) \qquad (i=1,2,3)\ , \nonumber
\ee 
where $\alpha_i=g_i^2/4\pi$ are the usual gauge couplings of $SU(3)_c\times SU(2)\times U(1)_Y$ at the messenger scale,  $x$ has been defined in eq.~(\ref{x}), and  
\begin{eqnarray*}
\label{g}
g(x)&=&\frac{1}{x^2}\left[(1+x)\log(1+x)+(1-x)\log(1-x)\right] \ , \\ 
g(x)&\simeq&1+\frac{x^2}{6}+\frac{x^4}{15}+\frac{x^6}{28} + \co(x^8)\ .
\end{eqnarray*}

On the other hand, the scalar masses arise from two-loop diagrams. Calculation of these graphs gives:
\be
m^2_{\tilde f} = 2\Lambda^2 N_5 \sum_{i=1}^3 C_i^{\tilde f}  \left(\df{\alpha_i}{4\pi}\right)^2f(x)\ , \nonumber
\ee
where $C_i$ are the corresponding quadratic Casimir operators [$C_N = (N^2-1)/(2N)$ for $SU(N)$] and
\begin{eqnarray*}
\label{f}
f(x)&=&\frac{1+x}{x^2}\left[\log(1+x)-2{\rm Li}_{2}\left(\frac{x}{1+x}\right)+\frac12{\rm Li}_{2}\left(\frac{2x}{1+x}\right)\right] \nonumber \\
&&+ (x\rightarrow-x) \ , \\ 
f(x)&\simeq&1+\frac{x^2}{36}-\frac{11x^4}{450}-\frac{319x^6}{11760} + \co(x^8) \ ,
\end{eqnarray*}
where ${\rm Li}_2$ is the dilogarithm (Spence's function).

\subsection{Models with radiatively generated $A$-terms}
\label{CPZappendix}
In these models, the presence of cubic operators in the superpotential, involving MSSM and messenger superfields, leads to trilinear couplings generated at one-loop level \cite{Evans:2011bea,Evans:2012hg}.  For the Evans and Shih model of eq.~(\ref{Evans}) the stop trilinear coupling reads
\be
A_t= -\frac{5\Lambda}{16\pi^2}\lambda^2 j(x), \nonumber
\ee
where
\begin{eqnarray*}
j(x)&=& \frac{1}{2x}\log\left(\frac{1+x}{1-x}\right) \ , \\
j(x)&\simeq& 1+\frac{x^2}{3}+\frac{x^4}{5}+\frac{x^6}{7}+\co(x^8).
\end{eqnarray*}

In addition, the following contributions to the scalar soft masses appear at one loop:
\begin{eqnarray*}
\label{ESsqmasses}
\delta m_{U_3}^2&=&-\frac{\Lambda^2}{48\pi^2}\lambda^2 x^2 h(x) \ , \\
\delta m_{H_u}^2&=&-\frac{\Lambda^2}{32\pi^2}\lambda^2 x^2 h(x) \ ,
\end{eqnarray*}
where 
\begin{eqnarray*}
\label{CPZh}
h(x)&=&\frac{3}{x^4}\left[(x-2)\log(1-x)-(2+x)\log(1+x)\right] \ , \\ 
h(x)&\simeq&1+\frac{4x^2}{5}+\frac{9x^4}{14}+\frac{8x^6}{15} + \co(x^8)\ .
\end{eqnarray*}

There are also two-loop contributions to the soft scalar masses that read
\begin{eqnarray*}
\label{EStwoloop}
\delta m_{Q_3}^2&=&-\frac{5\Lambda^2}{256\pi^4}\lambda^2 y_t^2 f_{h,2}(x) \ , \\
\delta m_{U_3}^2&=&\frac{\Lambda^2}{128\pi^4}\lambda^2 \left[6\lambda^2 + 2y_t^2-\frac{13}{15}g_1^2-3g_2^2-\frac{16}{3}g_3^2 \right] f_{h,2}(x) \ , \\
\delta m_{H_u}^2&=&\frac{\Lambda^2}{128\pi^4}\lambda^2 \left[9\lambda^2 + 3y_t^2-\frac{13}{10}g_1^2-\frac92g_2^2-8g_3^2 \right] f_{h,2}(x) \ , 
\end{eqnarray*}
where $y_t$ is the top Yukawa coupling, evaluated at $\Mm$, and $f_{h,2}(x)$ can be found in ref.~\cite{Jelinski:2015voa}.

Regarding the CPZ model of eq.~(\ref{extraW3}), 
the trilinear couplings for the stop and the sbottom read
\begin{eqnarray*}
\label{CPZtrilinear}
A_t= -\frac{3\Lambda}{16\pi^2}|\lambda_U|^2 y_t j(x) \ , \\
A_b= -\frac{\Lambda}{16\pi^2}|\lambda_U|^2 y_b j(x) \ ,
\end{eqnarray*}
where $y_t$ and $y_b$ are the top and bottom Yukawa couplings, respectively, evaluated at $\Mm$.

There are also one-loop contributions to the sfermions masses, given by 
\begin{eqnarray*}
\label{CPZsqmasses}
\delta m_{Q_3}^2&=&-\frac{\Lambda^2}{96\pi^2}|\lambda_U|^2 x^2 h(x) \ , \\
\delta m_{U_3}^2&=&-\frac{\Lambda^2}{48\pi^2}|\lambda_U|^2 x^2 h(x) \ ,
\end{eqnarray*}

Finally, the two-loop contributions to the soft scalar masses for $x \ll  1$ read
\begin{eqnarray*}
\label{extram2}
\delta m_{Q_3}^2&=&\frac{\Lambda^2}{256\pi^4}|\lambda_U|^2 \left[6|\lambda_U|^2 + 6y_t^2-\frac{13}{15}g_1^2-3g_2^2-\frac{16}{3}g_3^2 \right] \ , \\
\delta m_{U_3}^2&=&\frac{\Lambda^2}{128\pi^4}|\lambda_U|^2 \left[6|\lambda_U|^2 + 6y_t^2+y_b^2-\frac{13}{15}g_1^2-3g_2^2-\frac{16}{3}g_3^2 \right] \ , \\
\delta m_{D_3}^2&=&-\frac{\Lambda^2}{128\pi^4}|\lambda_U|^2 y_b^2 \ , \\
\delta m_{H_u}^2&=&-\frac{9\Lambda^2}{256\pi^4}|\lambda_U|^2 y_t^2 \ , \\
\delta m_{H_d}^2&=&-\frac{3\Lambda^2}{256\pi^4}|\lambda_U|^2 y_b^2 \ .
\end{eqnarray*}

\newpage

\FloatBarrier

\bibliographystyle{spphys}       
\bibliography{refs}

\end{document}